\documentclass[journal]{IEEEtran}
\usepackage{ifpdf}
\usepackage{cite}
\usepackage{url}

\usepackage[justification=centering]{caption}
\usepackage{amsmath}
\usepackage{multirow}
\usepackage{graphicx}
\usepackage{subfig}
\usepackage{epstopdf}
\usepackage{booktabs}
\usepackage{longtable}
\usepackage{multicol}
\usepackage{algorithm}
\usepackage{algorithmicx}
\usepackage{algpseudocode}
\usepackage{amssymb}
\usepackage{bbding}
\usepackage{bm}
\usepackage{colortbl}
\usepackage[table]{xcolor}
\usepackage{makecell}
\usepackage{amsfonts}
\begin{document}
\title{Reliable Distributed Computing for Metaverse: A Hierarchical Game-Theoretic Approach}
\author{Yuna Jiang, Jiawen Kang, Dusit Niyato,~\IEEEmembership{Fellow,~IEEE,} Xiaohu Ge,~\IEEEmembership{Senior Member,~IEEE,}

Zehui Xiong,~\IEEEmembership{Member,~IEEE,} Chunyan Miao,~\IEEEmembership{Senior Member,~IEEE,}
Xuemin(Sherman) Shen,~\IEEEmembership{Fellow,~IEEE}

\thanks{Yuna Jiang and Xiaohu Ge (Corresponding author) are with School of Electronic Information and Communications, Huazhong University of Science
and Technology, Wuhan 430074, Hubei, China and
Shenzhen Huazhong University of Science and Technology Research Institute, Shenzhen, 518063, Guangdong, China (e-mail: yunajiang@hust.edu.cn, xhge@mail.hust.edu.cn).

Jiawen Kang is with the School of Automation, Guangdong University of Technology, Guangzhou, 510006, Guangdong, China (e-mail: kavinkang@gdut.edu.cn).

Dusit Niyato and Chunyan Miao are with the School of Computer Science and Engineering, Nanyang Technological University, Singapore, 639798 (e-mail: DNIYATO@ntu.edu.sg, ascymiao@ntu.edu.sg).

Zehui Xiong is with the Pillar of Information Systems Technology and Design, Singapore University of Technology and Design, Singapore 487372 (e-mail: zehui\_xiong@sutd.edu.sg).

Xuemin (Sherman) Shen is with the Department of Electrical and Computer Engineering, University of Waterloo, Waterloo, ON N2L 3G1, Canada (e-mail: sshen@uwaterloo.ca).
}}

\markboth{}%
{Shell \MakeLowercase{\textit{et al.}}: Bare Demo of IEEEtran.cls for IEEE Journals}
\maketitle

\begin{abstract}
The metaverse is regarded as a new wave of technological transformation that provides a virtual space for people to interact through digital avatars. To achieve immersive user experiences in the metaverse, real-time rendering is the key technology. However, computing intensive tasks of real-time rendering from metaverse service providers cannot be processed efficiently on a single resource-limited mobile device. Alternatively, such mobile devices can offload the metaverse rendering tasks to other mobile devices by adopting the collaborative computing paradigm based on Coded Distributed Computing (CDC). Therefore, this paper introduces a hierarchical game-theoretic CDC framework for the metaverse services, especially for vehicular metaverse. In the framework, idle resources from vehicles, acting as CDC workers, are aggregated to handle intensive computation tasks in the vehicular metaverse. Specifically, in the upper layer, a miner coalition formation game is formulated based on a reputation metric to select reliable workers. To guarantee the reliable management of reputation values, the reputation values calculated based on the subjective logical model are maintained in a blockchain database.
In the lower layer, a Stackelberg game based incentive mechanism is considered to attract reliable workers selected in the upper layer to participate in rendering tasks.
The simulation results illustrate that the proposed framework is resistant to malicious workers. Compared with the best-effort worker selection scheme, the proposed scheme can improve the utility of metaverse service provider and average profit of CDC workers.
\end{abstract}


\IEEEpeerreviewmaketitle

\section{INTRODUCTION}
\subsection{Background and Motivations}
The blossom of emerging technologies, such as real-time rendering technologies, digital twin, artificial intelligence, 6G communications and blockchain, has promoted the proliferation of the metaverse \cite{a1}. The metaverse was first created in the science fiction named \emph{Snow Crash} \cite{a2}, which is a stereoscopic virtual space parallel to the physical world. Then, the successful broadcast of the famous film \emph{Ready Player One} raises the public's attention in the metaverse again \cite{a3}. 
To provide immersive experiences for people in the metaverse, real-time rendering technologies (e.g., extended reality and spatial sounding rendering) are considered to be the main interaction interfaces \cite{a4}. For the medium graphic and audio rendering services in the metaverse, the providers may not have dedicated computing resources. Besides, the intensive computation from medium metaverse services may be unbearable for resource-limited mobile devices \cite{a1}.
As such, the distributed collaborative computing has been adopted to solve computing-intensive tasks in the metaverse \cite{a5}.

For the distributed computing of metaverse services, multiple mobile devices work collaboratively to complete a large-scale rendering task. One of the main challenges of the distributed computing system for metaverse services is the straggler effects. The stragglers refer to the mobile devices whose computing speed is observably slower than average because of their limited computing resources or poor communication link, thus causing long latency and bad immersive experience for the metaverse, especially the metaverse interactive services (e.g., online games Minecraft and Roblox). Coded Distributed Computing (CDC) is a promising distributed computing solution to alleviate the straggler effects and guarantee fault-tolerance by aggregating the extra computing resources from mobile devices \cite{a6}. CDC introduces computing redundancy to the metaverse by code techniques, and metaverse service providers (MSPs) only need to collect the computing results of graphic and voice rendering tasks from a subset of workers. Thus CDC can significantly reduce the computation latency and improve the data processing reliability, equivalently alleviate the straggler effects. However, there are two critical problems that need to be addressed: i) The mobile devices may be unwilling to participate in metaverse services without a reasonable incentive; ii) Some mobile devices may even misbehave to damage MSPs' benefits, thus resulting in bad user experience. How to select reliable mobile devices and incentivize them to participate in metaverse services are still challenging.
\begin{figure}[!t]
	\begin{center}
		\includegraphics[width=0.45\textwidth]{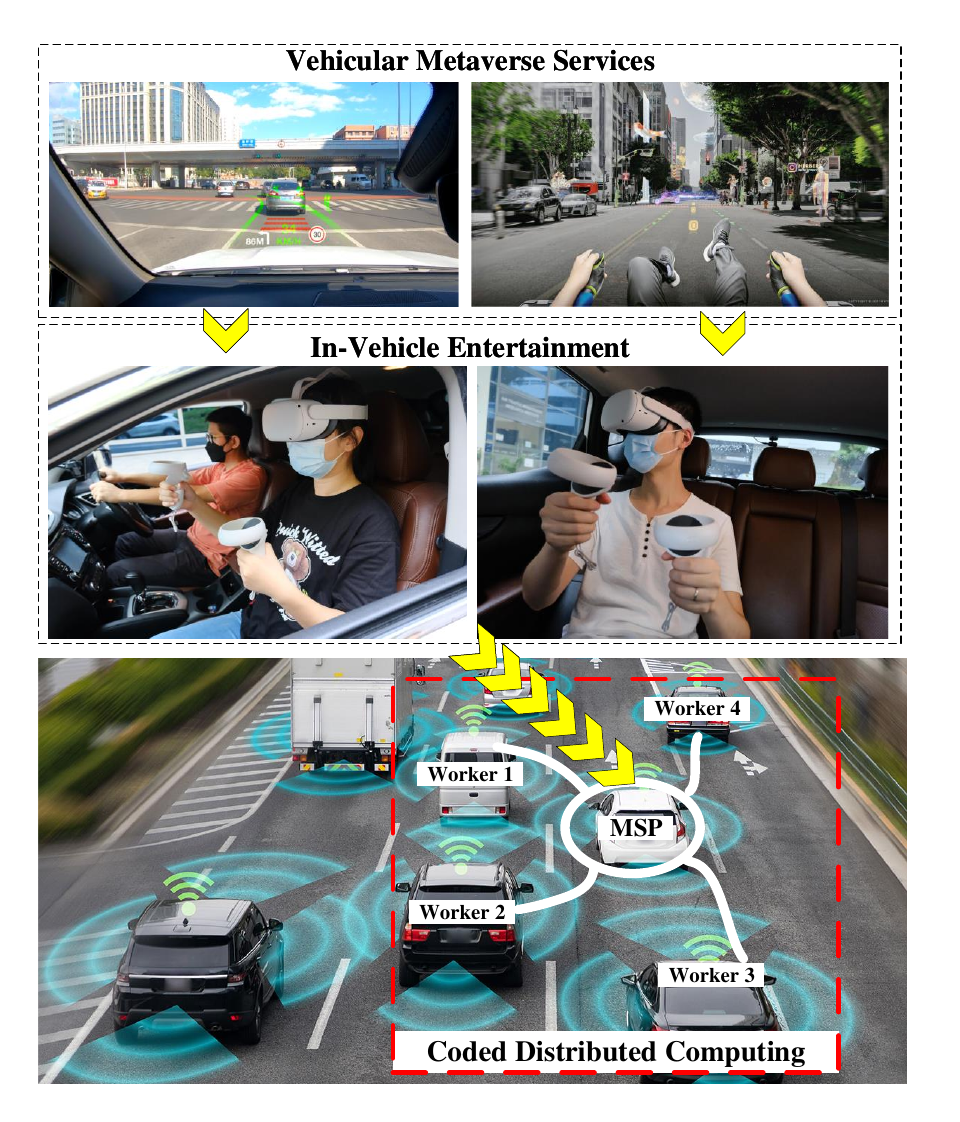}
	\end{center}
	\caption{An example of CDC in vehicular metaverse.}
	\label{Fig1}
\end{figure}

\subsection{CDC Use Case in Vehicular Metaverse}
The vehicular metaverse integrates extended reality technologies and real-time motion data seamlessly to blend virtual and real sapce for drivers and passengers in vehicles, which is an emerging in-vehicle entertainment segment for the automotive market. WayRay's Holograktor is striving to create the metaverse vehicle, where passengers and drivers are able to interact with a different reality during the vehicle ride. Figure 1 shows an example of CDC in the vehicular metaverse. With the vehicular metaverse services, passengers can entertain themselves with hybrid real and virtual environments, which may cause intensive computation for the resource-limited vehicles due to the graphic or audio rendering. In such condition, those vehicles can act as MSPs to cooperate with nearby vehicles that act as workers. However, this vehicular metaverse is vulnerable to stragglers due to not only the misbehavior, but also the highly mobile environments which can make some workers unable to complete computation tasks and return the results successfully.
Based on the CDC technology, MSPs can decompose and assign the rendering tasks to  workers, and MSPs only need to collect  a subset of workers' computing results, which is beneficial for the immersive experience of users in vehicles.

\subsection{Contributions}
In this paper, we mainly investigate the use case of CDC in vehicular metaverse, but the proposed scheme is not limited to the vehicular sector. We first adopt reputation metric to assess the reliability of workers in CDC to support vehicular metaverse services, and a stable coalition of reliable workers is formed based on reputation values. Moreover, we apply a hierarchical decision making structure composed of coalition formation and Stackelberg game to design an efficient incentive mechanism for reliable CDC in the vehicular metaverse. 
Blockchain is used to achieve distributed secure reputation management and to maintain worker participation records in the vehicular metaverse. As blockchain can manage the interactions among entities in the metaverse by a decentralized, tamper-proof and transparent manner \cite{metaverse_and_blockchain_review}.
We summarize the contributions of this paper as follows.
\begin{enumerate}
	\item We propose a novel reliable distributed collaborative computing framework for the metaverse based on CDC and blockchain technologies, which can support the immersive user experiences in the metaverse. Especially, we consider the use case in the vehicular metaverse.
	\item We adopt the reputation values to evaluate the reliability of workers. The miners who form coalitions are responsible for the calculation of workers' reputation values, and the reputation metric is the abstraction and aggregation of multiple factors that affect the quality of coalitions. 
	\item We propose a hierarchical game-theoretic approach to investigate the reliable and sustainable CDC scheme in the vehicular metaverse. In the upper layer, the coalition game is innovatively combined with the reputation metric to choose reliable CDC workers for MSPs. In the lower layer, the Stackelberg game is formulated to stimulate the reliable CDC workers to participate in metaverse services.
	\item Numerical results indicate that the proposed hierarchical game-theoretic scheme is resistant to the malicious workers. The MSP can obtain higher utility and CDC workers can obtain higher profits with the proposed scheme than those of the best-effort worker selection scheme.
\end{enumerate}

The strcture of this paper is shown as follows. Related works about CDC and metaverse services are introduced in Section II. The system model and problem formulations of the proposed reliable CDC scheme for the vehicular metaverse are presented in Section III. 
The worker selection process based on the reputation model and coalition game is presented in Section IV. Stackelberg game-based incentive mechanism is introduced in Section V. Section VI shows performance evaluation of reliable worker selection and CDC incentive mechanism. Section VII gives the conclusion of this paper.

\section{RELATED WORKS}
\subsection{Coded Distributed Computing }
The research on CDC mainly focuses on mitigating the straggler effects and minimizing the communication cost. Coding theories are the popular methods to solve above issues. In \cite{b12}, the authors explore the usage of coding theory to alleviate the straggler effect in matrix multiplication and reduce communication bottlenecks in data shuffling. The theoretical analysis proves that the coded schemes can obtain significant gains compared with uncoded schemes. In \cite{b13}, the authors study the distributed matrix multiplication problem in heterogeneous environments, where a coding framework is designed to accelerate distributed matrix multiplication with straggling devices.
In \cite{b14}, the authors consider that the straggler phenomenon of wireless distributed computing networks is caused by local computation and wireless transmission. Besides, the straggling factor for each device is derived. In \cite{b15}, a coded computing framework for federated learning is proposed, which uses structured coding redundancy to alleviate the straggler effects and speed up the training procedure. In \cite{b16}, the authors design the incentive mechanism for the CDC tasks by formulating a game-theoretic approach. In \cite{b17}, the authors design platforms' incentive mechanisms to encourage workers' participation in the coded machine learning. However, the existing works ignore the reliability of CDC workers when study CDC, which might negatively impact the performance of CDC. Besides, reliable CDC incentive mechanism for the metaverse has not been studied.

\subsection{Metaverse Services}
Metaverse-related research  is still in its infancy. There are some works studying the metaverse services in graphic or audio rendering, extended reality technologies etc. In \cite{a7}, the authors propose a vision of the metaverse native communications that includes encrypted address-based access model and blockchain.
In order to realize the audio/visual and virtual/reality congruence in metaverse services, the authors in \cite{a8} design the 6-degree-of-freedom interactive audio engines based on objects.
In \cite{a9}, the authors give a comprehensive survey on computational arts that blend virtual and physical environment in the metaverse. 
The authors in \cite{a10} design a brain-to-speech scheme for smart communication in the real world, which is also presented as a potential application in the metaverse. 
The authors in \cite{a11} propose an operating system for the metaverse based on extended reality, which integrates hardware, computer vision and extended reality specific network. 
In \cite{a12}, the authors propose a blockchain-based framework for the metaverse. The sharding scheme is used to improve the scalability of blockchain networks.
The authors in \cite{a13} give a survey about the Artifical Intelligence (AI) for the metaverse, and they especially explore AI-based methods that have potentials for the metaverse. The above works have not considered the metaverse services in vehicular networks. As metaverse might have a profound impact in the automotive field, a great deal of research needs to be done on the vehicular metaverse, including the distributed computing that is meaningful for the user immersive experience in the vehicular metaverse.

\begin{figure*}[htbp]
\begin{center}
\includegraphics[width=0.95\textwidth]{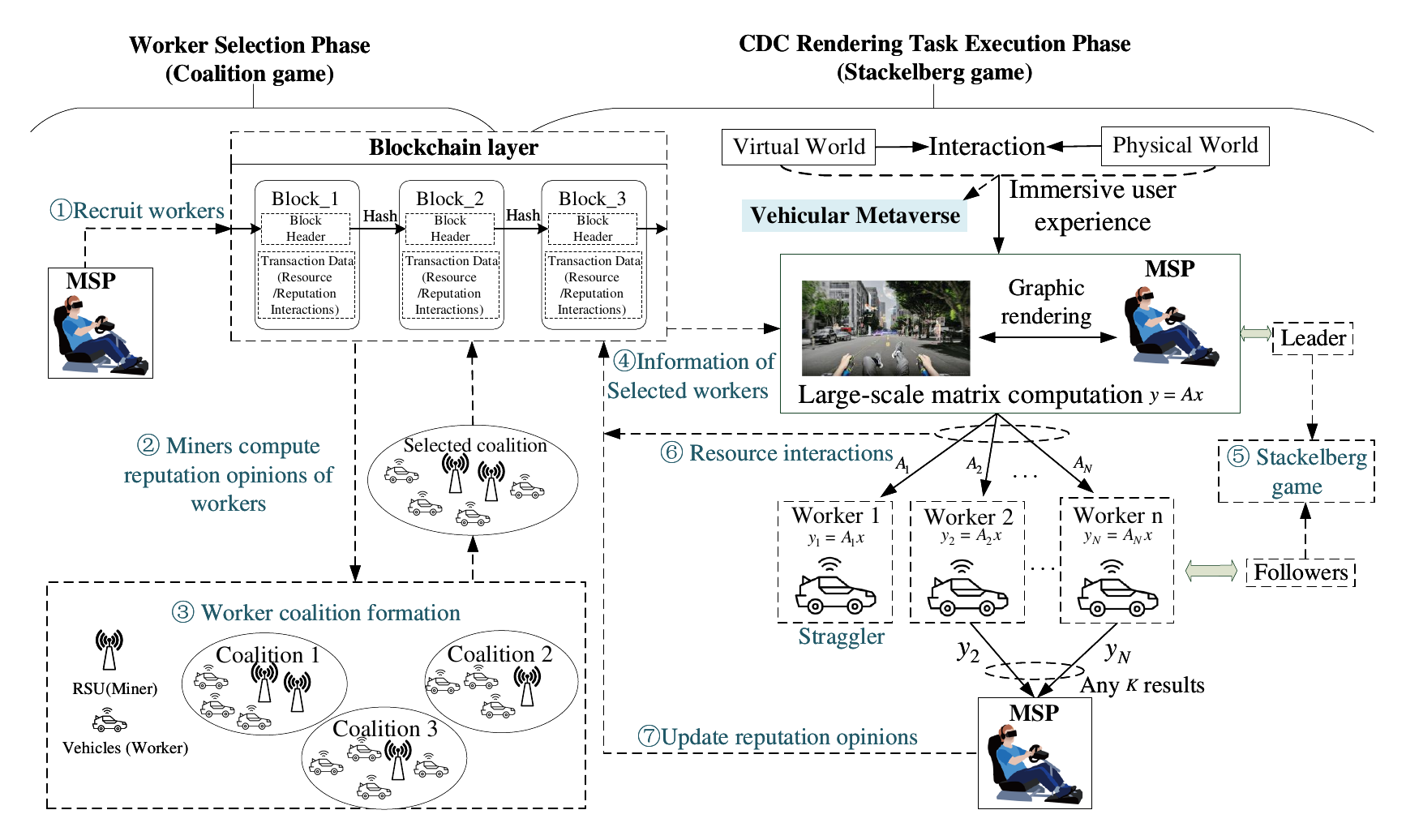}
\end{center}
\caption{System model: Worker selection and CDC rendering task execution in vehicular metaverse.}
\label{Fig2}
\end{figure*}
\section{SYSTEM MODEL AND PROBLEM FORMULATIONS}
\subsection{System Model}
 The vehicular metaverse mainly includes the virtual world, the physical world, and the interaction layer that connects the virtual and physical world. The immersive user experience is a significant part in the interaction layer \cite{a3}. To achieve the immersive user experiences, key technologies (e.g., graphic and spatial sound rendering) are adopted, which can easily lead to intensive matrix computation tasks to be processed by onboard units or devices in vehicles. Those vehicles can choose to execute rendering tasks by collaborating with other vehicles within a limited communication distance. In the metaverse, vehicles that distribute the large-scale matrix computation tasks (e.g., projection and shadow mapping in graphic rendering \cite{d2}) are called MSPs. The vehicles that execute the computation tasks of metaverse services are called workers. In order to realize the reliable distributed computing, the MSP first selects workers based on reputation values, and then allocates the rendering tasks to workers based on the CDC technology. Blockchain is used to manage reputation values and to record worker participation, realizing the decentralized CDC in the vehicular metaverse. RodeSide Units (RSUs) with sufficient computation and storage resources in vehicular networks act as miners to perform consensus algorithm. In addition, miners  are stimulated to join the reputation calculation since they are more trustable entities than vehicles. Miners that have contributions to the reliable worker selection will be rewarded
 by blockchain (e.g., receiving some tokens or obtaining resource rewards).

Figure 2 shows the whole CDC process in the vehicular metaverse including the worker selection phase and CDC rendering task execution phase. In the worker selection phase, the MSPs select workers based on workers' reputation values recorded on blockchain. In the CDC task execution phase, MSPs distribute the rendering tasks to the selected workers. The workers process the computation tasks and transmit the final results to MSPs. Then MSPs update workers' reputation values and record the updated values on blockchain. Besides, all the resource interactions (e.g, workers contribute their computing resource to MSPs) among MSPs and workers are recorded on blockchain. The details about the CDC process are introduced as follows.

\emph{Step 1}: The MSP sends a transaction to blockchain to recruit workers that are willing to serve for the rendering tasks in the vehicular metaverse. The transaction contains the requirements of the MSP, such as the reputation threshold $T_{th}^{com}$ of workers.

\emph{Step 2}: The workers that are willing to help the MSP with metaverse services send transactions to respond to the MSP. The worker set is denoted as $\mathbb{W} = \left\{{1,2, \ldots ,w, \ldots ,W} \right\}$. In the metaverse, users have many perception dimensions in immersive experience due to computations of graphic rendering and sound generation. Such perception can be qualitative and the subjective logical model is suitable.
In order to select trustable workers, the blockchain needs to obtain the compositive reputation opinions of workers based on the subjective logical model \cite{compositive_reputation}. Miners are stimulated to compute workers' compositive reputation opinions. The miners that are
selected by blockchain based on their contribution to workers’
reputation calculation will be rewarded by blockchain. The set of miners is denoted as $\mathbb{M}= \left\{ {1,2, \ldots ,m,\ldots ,M} \right\}$. 
Miners have no idea about the reliability of workers before they compute the reputation values of workers.
Each miner randomly selects several workers and computes workers' compositive reputation opinions. The number of workers selected by miner $m$ is denoted as $\left| {{{\mathbb {W}}_m}} \right|$. The workers whose compositive reputation opinions are lower than $T_{th}^{com}$ are discarded by miners, and $T_{th}^{com}$ is set based on MSPs' service requirements \cite{compositive_reputation}. We assume that the number of discarded workers is much smaller than those of workers that are not abandoned by miners.
The set of workers selected by miner $m$ is denoted as ${\mathbb{W}_m} = \left\{ {{{\cal W}_{m,1}},{{\cal W}_{m,2}}, \ldots ,{{\cal W}_{m,w}}, \ldots {{\cal W}_{m,\left| {{\mathbb{W}_m}} \right|}}} \right\}$, where ${{{\cal W}_{m,w}}}$ is the identity number of the $w$-th worker selected by miner $m$.

\emph{Step 3}: In order to increase a likelihood to be rewarded by blockchain,  miners that obtain workers' compositive reputation opinions form coalitions. The coalition members in each coalition contain miners and their corresponding worker sets.
When the coalition is selected by blockchain, the compositive reputation values of workers in the selected coalition will be recorded on blockchain by miners. Another benefit for the coalition formation is that multiple reliable workers evaluated
by their reputation can offer larger computational resources to the MSP.

\emph{Step 4}: The blockchain returns selected workers' information, including the
location information and reputation opinions, to the MSP.

\emph{Step 5}: The MSP distributes the rendering tasks to the selected workers based on Stackelberg game. In the game, the MSP acts as the leader to determine the reward strategy, and workers act as followers to adjust computing speed strategy. Specifically, we consider that MSP ${P_i}$ needs to render 2D images in RGB color into the needed Field of Views (FoV). The number of pixels is denoted as ${n_p}$, and the resolution of the required FoV is ${n_p} \times {n_p}$. With the CDC technology, the MSP allocates the rendering task to workers. The Maximum Distance Separable (MDS) code is adopted to alleviate the straggler effects in the distributed computing \cite{a6}. The rendering task of MSP ${P_i}$ is expressed as  ${y_i} ={A_i}{x_i}$, and ${{A_i} \in {\mathbb{R}^{(3\times2\times{n_p}) \times {n_p}}}} $, where \textquotedblleft 3" is the number of colors in RGB model, which contains red, green and blue colors, \textquotedblleft 2" is the number of viewpoints \cite{JSAC_wireless_VR_Xiaonan}. MSP ${P_i}$ first divides the rendering matrix ${A_i}$ into $K$ equal-sized submatrices in
${\mathbb{R}^{\frac{{(3\times2\times{n_p})}}{K} \times {n_p}}}$. Then, based on $\left( {N,K}
\right)$ MDS code, the MSP gets $N$ encoded submatrices with the same size $\frac{{(3\times2\times{n_p})}}{K} \times {n_p}$.
Each submatrix is allocated to a worker. MSP ${P_i}$ can reconstruct the final rendering
result when receiving the results from any $K$($K < N$) workers, which can mitigate straggler effects.

\emph{Step 6}: The resource interactions between the MSP and workers are recorded on blockchain. Then, the workers can receive the deserved reward from the MSP.

\emph{Step 7}: After finishing the CDC rendering tasks, the MSP updates the local reputation opinions of workers on blockchain.

To enable immersive vehicular metaverse experience and realize the CDC rendering task execution efficiently, the MSP can select workers with the help of blockchain in advance. When the MSP has distributed computing requirements, the MSP can distribute the rendering task to workers in time.
\subsection{Transmission Model }
We consider that the CDC rendering task execution mainly contains computing the tasks and transmitting the computing results to the MSP. Here we establish the transmission model that vehicle workers transmit the rendering task computation results to the MSP, where we adopt the orthogonal frequency division multiple access (OFDMA) scheme. MSP ${P_i}$ uses a set of resource blocks for the wireless transmission, and there is no interference between workers, this assumption is typically to simply the analysis, but the model can still be straightforwardly extended with interference.
We also consider the effects of mobility on channel gain between the MSP ${P_i}$ and workers. To estimate an accurate mobile channel gain, the channel state information of mobile link is broadcast to workers with a period of ${T_c}$. The channel states of the wireless mobile links in the previous and current time interval are denoted as $g$ and $\hat g$, respectively. The relationship between $g$ and $\hat g$ is $g = \varpi \hat g + \kappa$
, where $\varpi  = {J_0}\left( {2\pi {f_D}{T_c}} \right) (0 < \varpi  < 1)$ is the channel correlation coefficient. ${J_0}\left(  \bullet  \right)$ is the zero-order Bessel function. ${f_D}$ is the maximum Doppler frequency and is expressed as ${f_D} = {{\left| {\Delta v} \right|{f_c}} \mathord{\left/{\vphantom {{\left| {\Delta v} \right|{f_c}} {3 \times {{10}^8}}}} \right.\kern-\nulldelimiterspace} {3 \times {{10}^8}}}$. ${\left| {\Delta v} \right|}$ is the relative vehicle speed between the MSP and workers. ${{f_c}}$ is the carrier frequency at 5.9GHz. $\kappa$ is the channel discrepancy coefficient that follows $\mathcal{CN}\left( {0,1 - {\varpi ^2}} \right)$ \cite{ vehicle_mobility_1}, \cite{ vehicle_mobility_2}. Then the channel gain model between MSP ${P_i}$ and worker $w$ is expressed as
\begin{equation}
\mathcal H_{iw} = G_{iw}\left( {{{\left( {\varpi _{iw}\hat g_{iw}} \right)}^2} + {{\left| {\kappa _{iw}} \right|}^2}} \right),
\end{equation}
where $G_{iw}$ is the large-scale fading effect between MSP ${P_i}$ and worker $w$. 
The signal-to-noise between MSP ${P_i}$ and worker $w$ is expressed as 
\begin{equation}
{\eta _{iw}} = \frac{{{p_w}\mathcal H_{iw}}}{{{{\tilde \sigma }^2} }},
\end{equation}
where ${p_w}$ is the transmitting power of worker $w$, ${{{\tilde \sigma }^2}}$ is the variance of the Gaussian noise. The data rate between worker $w$ and MSP ${P_i}$ is expressed as 
\begin{equation}
{c_{iw}} = B{\log _2}\left( {1 + {\eta _{iw}}} \right).
\end{equation}
The communication delay of worker $w$ is expressed as
\begin{equation}
T_w^{com - u}{\rm{  = }}\frac{{{s_w}}}{{{c_{iw}}}},
\end{equation}
where $s_w$ is the packet size of computation results transmitted by worker $w$.

\subsection{Computation Model}
The computation time of each worker follows a 2-parameter shifted
exponential distribution \cite{b12},\cite{b13}. Then, the
cumulative distribution function of the time $T_w^{cmp}$ that worker $w$
finishes the rendering task is expressed as

\begin{equation}
	\Pr \left( {T_w^{cmp} \le t} \right) = 1 - {e^{ - {\mu _w}\left(
			{\frac{t}{{{l}}} - {a_w}} \right)}},{\rm{  }}\forall {\rm{t}} \ge {a_w}{l},
\end{equation}
where ${\mu _w}$ is the average computation speed of worker $w$, ${a_w}$ is the start-up time to begin the rendering task computation, ${l}$ is the amount of rendering task allocated to each worker, and ${{l = 3 \times 2 \times {n_p}} \mathord{\left/{\vphantom {{l = 3 \times 2 \times {n_p}} K}} \right.
\kern-\nulldelimiterspace} K}$. The probability density function of the above distribution is 

\begin{equation}
	f\left( t \right) = \frac{{{\mu _w}}}{{{l}}}{e^{ - {\mu _w}\left(
			{\frac{t}{{{l}}} - {a_w}} \right)}}.{\rm{ }}
\end{equation}
The average computation time for worker $w$ is expressed as
\begin{equation}
	\begin{aligned}
		\mathbb{E}\left(T_{w}^{c m p}\right) &=\int_{0}^{+\infty} t \frac{\mu_{w}}{l} e^{-\mu_{w}\left(\frac{t}{l}-{a_w}\right)} d t, \\
		&=\frac{l e^{{a_w} \mu_{w}}}{\mu_{w}}.
	\end{aligned}
\end{equation}
The average rendering task execution time of worker $w$ is expressed as
\begin{equation}
	\mathbb{E}\left( {{T_w}} \right) = \mathbb{E}\left( {T_w^{com - u}} \right) + \mathbb{E}\left( {T_w^{cmp}}
	\right){\rm{ }}.
\end{equation}

\subsection{Hierarchical Game-theoretic CDC Framework for Vehicular Metaverse}
\begin{figure}[!h]
\begin{center}
\includegraphics[width=0.5\textwidth]{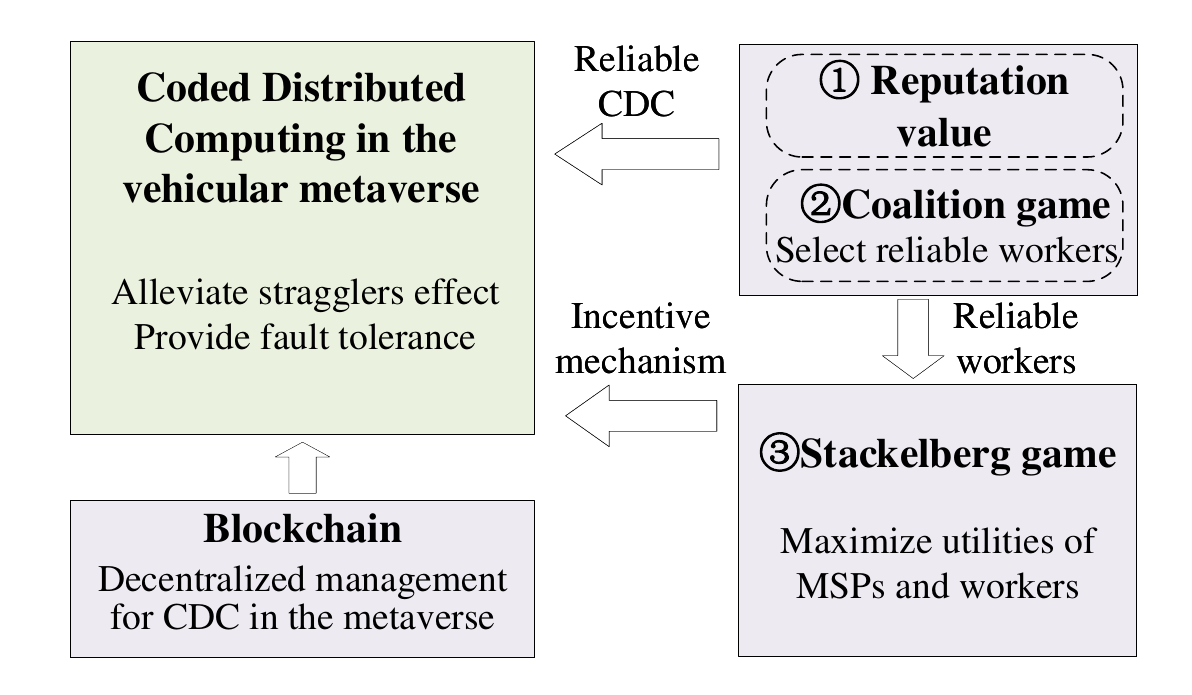}
\end{center}
\caption{Hierarchical game-theoretic CDC framework for blockchain assisted vehicular metaverse.}
\label{Fig3}
\end{figure}
The outline of the proposed hierarchical game is shown in Fig. 3. Blockchain is utilized to achieve the decentralized management for CDC in the vehicular metaverse. In order to realize reliable and sustainable CDC in the vehicular metaverse, we adopt a hierarchical game-theoretic approach based on coalition formation and Stackelberg game. The coalition-Stackelberg game makes the proposed framework suitable for reliable CDC in the vehicular metaverse. The description of the hierarchical game-theoretic approach is as follows:

1) Upper layer: In the upper layer, the coalition formation game is adopted to investigate the cooperative interactions among miners that contribute to the computation of workers' compositive reputation values. The miners compute workers' compositive reputation values based on reputation opinions stored on blockchain. Then the miners form coalitions by maximizing the coalition utility that includes workers' compositive reputation values and cost. In order to be selected by blockchain, each miner prefers to form coalitions with other miners to obtain higher coalition utility. Besides, considering the coalition formation cost, the grand coalition may not be stable. This is due to the fact that a high communication overhead discourage all the miners to participate and act as a coalition. The coalition with the highest coalition utility will be selected by blockchain, and the workers in the selected coalition will execute CDC rendering tasks.

2) Lower layer: In the lower layer, the Stackelberg game is used to incentivize workers selected in the upper layer to allocate more computing resource to the MSP's CDC rendering tasks. In the Stackelberg game, the MSP acts as a leader to adjust the reward to workers who finish the computing tasks in time, the workers act as followers to adjust the average computing speed. The strategies of workers also affect their local reputation values updated by the MSP.

3) Interactions between upper and lower layers: In the upper layer, the miners who compute workers' compositive reputation values form coalitions. The coalition with the highest coalition utility will be selected by blockchain, and the workers in the selected coalition will execute the CDC rendering tasks published by the MSP. Specifically, without forming coalitions, the MSP may not be able to identify the workers that are reliable, which may degrade the utility of the MSP. In order to motivate the selected workers to contribute their computing resource to the CDC rendering tasks and guarantee the highest utility of the MSP, the Stackelberg game is used in the lower layer. The MSP decides reward strategy, and the selected coalitions of workers, which are formed in the upper layer, contribute their computing resource to maximize their utilities. Besides, in order to increase probabilities to be selected in the next round of coalition formation game, the workers should also consider their local reputation values when adjusting the computing resource strategy. The hierarchical game theoretic approach ensures that the CDC rendering tasks can be executed by reliable workers and maximizing the MSP's utility. Meanwhile, the selected reliable workers are incentivized to contribute their computing resources to the CDC rendering tasks from vehicular metaverse services.
	
\section{RELIABLE WORKER SELECTION FOR VEHICULAR METAVERSE}
In the worker selection phase, we mainly focus on two problems. The first one is how to obtain the overall reputation opinions from the interaction histories between MSPs and workers. Subjective logic is a popular tool to model the reliability of entities in the vehicular metaverse, as it quantifies belief, disbelief and uncertainty \cite{b20}. Based on the subjective logic model, both direct and indirect reputations from other MSPs are combined to derive the compositive reputation opinions of workers. The second problem is how to select reliable workers based on the reputation opinions of workers in a distributed way. With the subjective logic model, the reputation calculation of workers may be not easy for the MSPs, as the MSPs need to search all the local and recommended reputation opinions of workers on blockchain, and then compute the workers' compositive reputation opinions. We motivate
miners in blockchain to calculate workers' reputation opinions. Each miner in $\mathbb{M}$ selects several workers from $ \mathbb{W}$ to compute the reputation opinions. The miners compute the selected workers' compositive reputation opinions based on the stored direct reputation opinions and subjective logic model. The workers whose
compositive reputation opinions are lower than the reputation threshold
$T_{th}^{com}$ are discarded by miners. Then, miners form coalitions to increase their chance to be rewarded by blockchain. The blockchain selects the miner coalition with the highest coalition utility. 
In this section, the subjective logic model is introduced, then the miner coalition formation game is formulated.

\subsection{Subjective Logic Model-Based Reputation Calculation}
\subsubsection{Local opinions for subjective logic}
The MSPs' direct reputation opinions to all workers are recorded on blockchain. MSP ${P_i}$ sends a transaction to blockchain to recruit workers. Then, the workers that are willing to help MSP ${P_i}$ send transactions to respond to MSP ${P_i}$.
Miners compute the workers' compositive reputation opinions by combining the direct reputation opinions updated by MSP ${P_i}$, and those updated by other MSPs on blockchain. MSP ${P_i}$'s direct reputation opinions are considered to be local reputation opinions, and other MSPs' direct reputation opinions are considered to be recommended reputation opinions.

The local opinion of MSP ${P_i}$ to worker $w$ in the subjective logic model is expressed as a vector $R_{i \to w}^{local} = \left\{ {b_{i \to w}^{local},d_{i \to w}^{local},u_{i \to w}^{local}} \right\}$, where $b_{i \to w}^{local}$ represents belief, $d_{i \to w}^{local}$ represents disbelief and $u_{i \to w}^{local}$
represents uncertainty. Here ${b_{i \to w}^{local}}$, ${d_{i \to w}^{local}}$, ${u_{i \to w}^{local}}$ $ \in\left[ {0,1} \right]$, and ${b_{i \to w}^{local}} + {d_{i \to w}^{local}} + {u_{i \to w}^{local}} = 1$. Based on the subjective logic models in \cite{c1}, ${b_{i \to w}^{local}}$, ${d_{i
\to w}^{local}}$ and ${u_{i \to w}^{local}}$ are represented as follows:
\begin{equation}
\left\{ \begin{array}{l}
b_{i \to w}^{local} = \frac{{{\sigma _1}{p_{i \to w}}}}{{{\sigma _1}{p_{i \to w}} + {\sigma _2}{q_{i \to w}}+2}},\\
d_{i \to w}^{local} = \frac{{{\sigma _2}{q_{i \to w}}}}{{{\sigma _1}{p_{i \to w}} + {\sigma _2}{q_{i \to w}}+2}},\\
u_{i \to w}^{local} = \frac2{{{\sigma _1}{p_{i \to w}} + {\sigma _2}{q_{i \to w}}+2}},
\end{array} \right.
\end{equation}
where ${p_{i \to w}}$ and ${q_{i \to w}}$ are the numbers of positive and negative interactions between MSP ${P_i}$ and worker $w$, respectively. The MSP regards the resource interaction between itself and a worker as a positive event if the worker is not a straggler in the CDC, and returns the effective computation results that is important for enhancing immersive experience of the vehicular metaverse services (e.g., graph resolution or audio playback resolution). 
The weights of positive and negative interaction are denoted as
${\sigma _1}$ and ${\sigma _2}$, respectively, and $0 < {\sigma _2} < {\sigma _1} < 1$. The local reputation value ${T_{i \to w}^{local}}$ is expressed as

\begin{equation}
{T_{i \to w}^{local}} = {b_{i \to w}^{local}} + \gamma {u_{i \to w}^{local}},
\end{equation}
where $\gamma  \in \left[ {0,1} \right]$ is the effective coefficient of
uncertainty on the reputation of worker $w$.
\subsubsection{Recommended opinions}
Apart from MSP ${P_i}$'s local reputation opinions, the miners also need to search for selected workers' direct reputation opinions updated by other MSPs on
blockchain to obtain the recommended reputation opinions. Suppose that
miner $m$ receives a number of $\Re $ recommended opinions of $w$ on blockchain. $\Re $ is also the number of recommenders. For recommender
$r \in \Re $, the weight factor ${\omega _r}$ is expressed as \cite{b23}

\begin{equation}
{\omega _r} = \frac{{{b_{i \to r}} \times \left( {b_{r \to w}^{local} + d_{r \to w}^{local}} \right)}}{{\sum\limits_{r \in \Re } {{b_{i \to r}} \times \left( {b_{r \to w}^{local} + d_{r \to w}^{local}} \right)} }},
\end{equation}
where $b_{r \to w}^{local} + d_{r \to w}^{local}$ represents the familiarity value between
recommender $r$ and worker $w$. The higher familiarity value means that the recommended opinions of recommender $r$ on worker $w$ is more convincing. ${b_{i \to r}}$ is the social tie strength between MSP ${P_i}$ and the recommender. When MSP ${P_i}$ and the recommender have the same worker set, we consider that the strength of social ties between MSP ${P_i}$ and the recommender is strong \cite{c2}. The sets of workers that have provided computing services for MSP ${P_i}$ and recommender $r$ are denoted as $\Gamma \left( {{P_i}} \right)$ and $\Gamma \left( r \right)$, respectively. Then, ${b_{i \to r}}$ is expressed as \cite{c2}
\begin{equation}
{b_{i \to r}} = \frac{{\left| {\Gamma \left( {{P_i}} \right) \cap \Gamma \left( r \right)} \right|}}{{\left| {\Gamma \left( {{P_i}} \right) \cup \Gamma \left( r \right)} \right|}}.
\end{equation}
The overall recommended reputation opinion of $w$ is denoted as $R_{i \to
w}^{rec} = \left\{ {b_{i \to w}^{rec},d_{i \to w}^{rec},u_{i \to w}^{rec}}
\right\}$. The recommended reputation opinion of worker $w$ is the combination of all the recommenders' local opinions with their weight ${\omega _r}$, $r \in \Re $. $b_{i \to w}^{rec}$, $d_{i \to w}^{rec}$ and $u_{i \to w}^{rec}$ are represented as follows:

\begin{equation}
\left\{ \begin{array}{l}
b_{i \to w}^{rec} = \sum\nolimits_{r \in \Re } {{\omega _r}b_{r \to w}^{local}} ,\\
d_{i \to w}^{rec} = \sum\nolimits_{r \in \Re } {{\omega _r}d_{r \to w}^{local},} \\
u_{i \to w}^{rec} = \sum\nolimits_{r \in \Re } {{\omega _r}u{{_{r \to w}^{local}}_.}}
\end{array} \right.
\end{equation}
\subsubsection{Combining local opinions with recommend opinions}
Based on the local opinion and recommended opinions from other MSPs.
Miner $m$ can obtain the final compositive reputation opinion of MSP ${P_i}$
to worker $w$. The compositive reputation opinion of MSP ${P_i}$ to
worker $w$ is represented as $R_{i \to w}^{com} = \left\{ {b_{i \to w}^{com},d_{i
\to w}^{com},u_{i \to w}^{com}} \right\}$.
According to \cite{compositive_reputation},
 $b_{i \to w}^{com}$, $d_{i \to
w}^{com}$, $u_{i \to w}^{com}$ are expressed as follows:

\begin{equation}
\left\{ {\begin{array}{*{20}{l}}
{b_{i \to w}^{com} = \frac{{b_{i \to w}^{local}u_{i \to w}^{rec} + b_{i \to w}^{rec}u_{i \to w}^{local}}}{{u_{i \to w}^{local} + u_{i \to w}^{rec} - u_{i \to w}^{rec}u_{i \to w}^{local}}},}\\
{d_{i \to w}^{com} = \frac{{d_{i \to w}^{local}u_{i \to w}^{rec} + d_{i \to w}^{rec}u_{i \to w}^{local}}}{{u_{i \to w}^{local} + u_{i \to w}^{rec} - u_{i \to w}^{rec}u_{i \to w}^{local}}},}\\
{u_{i \to w}^{com} = \frac{{u_{i \to w}^{rec}u_{i \to w}^{local}}}{{u_{i \to w}^{local} + u_{i \to w}^{rec} - u_{i \to w}^{rec}u_{i \to w}^{local}}}.}
\end{array}} \right.
\end{equation}
The compositive reputation value of MSP ${P_i}$ to worker $w$ is
expressed as
\begin{equation}
T_{i \to w}^{com} = b_{i \to w}^{com} + \gamma u_{i \to w}^{com}.
\end{equation}
For the initialization of workers' reputation values, the numbers of positive and negative interactions between MSPs and workers are set as 0. Then the initial reputation values of workers can be obtained based on the above model.
According to the compositive reputation opinions, MSPs can recruit more workers with high reputation values, promoting the finish of CDC rendering tasks.
After finishing a computing task, the MSP updates the selected workers' direct
reputation opinions on blockchain.

\subsection{Coalition Formation Game Formulations}
In the proposed model, miners cooperate to select reliable workers for vehicular metaverse services. To achieve suitable cooperative
strategies for miners, the coalition game theory is used. The combination of coalition game and reputation metric makes the coalition more multi-dimensional from the decision making perspective. The formulated model could be suitable than classical coalition formation game that relies only on a single value of utility, which may not be known precisely in reality. The cooperative worker selection problem among miners is modeled as a coalition formation game with non-transferable utility (NTU), which means that the value or the utility of a coalition cannot randomly be divided between the members in the coalitions \cite{b24}. The NTU coalition game is modeled as $ \mathbb{G}= \left\{ {\mathbb{M},u}
\right\}$, and $u$ is the utility function. The
blockchain returns the compositive reputation opinions of the workers in
the coalition with the highest utility.
We denote the coalition of miners as \textbf{ }${{\cal G}_o} \in \mathbb{M} $, and $o$ is the index of the coalition \cite{b25}.
In the following, we give the definition of coalition partition for the coalition formation game.

\textbf{Definition 1: }A group of mutually disjoint coalitions in $\mathbb{M}$ is represented as $\Pi  = \left\{ {{{\cal G}_1},{{\cal G}_2},
\ldots ,{{\cal G}_o}, \ldots ,{{\cal G}_O}} \right\}$, where ${{\cal G}_o} \cap
{{\cal G}_{o'}} = \phi $ for $o \ne o'$, $O$ is the number of coalitions.
If the group covers all the players in $\mathbb{M}$, i.e., $ \cup _{o = 1}^O{{\cal G}_o} =\mathbb{M}
$, the group is called a coalition partition of $\mathbb{M}$ \cite{b26}.

The worker set for the coalition ${{\cal G}_o}$ is denoted as ${ \cup _{m \in {{\cal G}_o}}}{\mathbb{W}_m}$, and the number of workers in ${ \cup _{m \in {{\cal G}_o}}}{\mathbb{W}_m}$ is denoted as $\left| {{ \cup _{m \in {{\cal G}_o}}}{\mathbb{W}_m}} \right|$. For the calculation of workers' reputation opinions, the contribution value of the coalition ${{\cal G}_o}$ is expressed as
\begin{equation}
Q\left( {{{\cal G}_o}} \right) = {\rho _c}\frac{{\left| {{ \cup _{m \in {{\cal G}_o}}}{\mathbb{W}_m}} \right|}}{W} + {\rho _r}\frac{{\sum\nolimits_{m \in {{\cal G}_o}} {\sum\nolimits_{w = 1}^{\left| {{\mathbb{W}_m}} \right|} {T_{i \to w}^{com}} } }}{{\left| {{ \cup _{m \in {{\cal G}_o}}}{\mathbb{W}_m}} \right|}},
\end{equation}
where ${{\left| {{ \cup _{m \in {{\cal G}_o}}}{\mathbb{W}_m}} \right|} \mathord{\left/ {\vphantom {{\left| {{ \cup _{m \in {{\cal G}_o}}}{_m}} \right|} W}} \right. \kern-\nulldelimiterspace} W}$ is the percent of workers that the coalition ${{\cal G}_o}$ has computed, and the second term ${{\sum\nolimits_{m \in {{\cal G}_o}} {\sum\nolimits_{w = 1}^{\left| {{\mathbb{W}_m}} \right|} {T_{i \to w}^{com}} } } \mathord{\left/{\vphantom {{\sum\nolimits_{m \in {{\cal G}_o}} {\sum\nolimits_{w = 1}^{\left| {{\mathbb{w}_m}} \right|} {T_{i \to w}^{com}} } } {\left| {{ \cup _{m \in {{\cal G}_o}}}{_m}} \right|}}} \right.
\kern-\nulldelimiterspace} {\left| {{ \cup _{m \in {{\cal G}_o}}}{\mathbb{W}_m}} \right|}}$ is the average reputation values of workers in the coalition ${{\cal G}_o}$. ${\rho _c}$ and ${\rho _r}$ are the coefficients that maintain the importance of the above two parts on the contribution value.

The communication cost among miners increases with the growth of the number of miners in the coalition, because forming a miner coalition requires negotiation and information exchange that may bring the cost and reduce the gains from forming the coalition \cite{b27}. For the coalition ${{\cal G}_o}$, the communication cost ${\cal C}\left( {{{\cal G}_o}} \right)$ should reflect the negotiation and information exchange overhead, which are determined by the number of miners in the coalition ${{\cal G}_o}$. The communication cost ${\cal C}\left( {{{\cal G}_o}} \right)$ should satisfy the following requirements. Firstly, the communication cost ${\cal C}\left( {{{\cal G}_o}} \right)$ increases monotonically with respect to
$\left|\mathbb{M}{{_o}} \right|$ that is the number of miners in the coalition ${{\cal G}_o}$. Secondly, the slope of ${\cal C}\left( {{{\cal G}_o}}
\right)$ becomes steeper with the increase of $\left|\mathbb{M}{{_o}} \right|$. In
order to satisfy the above requirements, the logarithmic barrier penalty function is
adopted to model the coalition cost \cite{b28},\cite{b29}. The communication cost ${\cal C}\left( {{{\cal G}_o}} \right)$ is expressed as
\begin{equation}
\mathcal{C}\left(\mathcal{G}_{o}\right)=\left\{\begin{array}{ll}
-\log \left(1-\left(\frac{\left|\mathbb{M}_{o}\right|-\tilde o}{M}\right)^{2}\right), & \text { if }\left|\mathbb{M}_{o}\right| \geq 2, \\
0, & \text { otherwise },
\end{array}\right.
\end{equation}
where $\tilde o$ is used to avoid an infinite value of ${\cal C}\left( {{{\cal G}_o}}
\right)$ when $\left|\mathbb{M} {{_o}} \right| = M$, and $\tilde o = 0.1$ \cite{b28}. Then, the coalition utility of the coalition ${{\cal G}_o}$ is expressed as
\begin{equation}
u\left( {{{\cal G}_o}} \right) = Q\left( {{{\cal G}_o}} \right) - \delta {\cal C}\left( {{{\cal G}_o}} \right),
\end{equation}
where $\delta$ is the communication cost coefficient.

For the coalition ${{\cal G}_o}$, whether the miners in ${{\cal G}_o}$ can be rewarded by blockchain depends on the coalition utility $u\left( {{{\cal G}_o}} \right)$. We notice that, in the proposed coalition game, the coalition utility $u\left( {{{\cal G}_o}} \right)$ is not divisible among the miners, and the utility of each miner in the coalition ${{\cal G}_o}$ is equal to $u\left( {{{\cal G}_o}} \right)$, instead of a fraction of $u\left( {{{\cal G}_o}} \right)$. Hence, the proposed coalition formation game has non-transferable utility.
Each miner can choose the suitable coalition based on the received utility. The definitation of preference order is given as follows.

\textbf{Definition 2: } A preference operator $\triangleright$ is adopted to compare ${\Pi _1} = \left\{ {{\cal G}_1^1, \ldots ,{\cal G}_O^1} \right\}$ and ${\Pi _2} = \left\{ {{\cal G}_1^2, \ldots ,{\cal G}_{O'}^2} \right\}$ that are partitions of the same subset $\mathbb{A} \subseteq \mathbb{M}$ (i.e., the players in ${\Pi _1}$ and ${\Pi _2}$ are the same). Then, ${\Pi _1} \triangleright {\Pi _2}$ implies that
partition ${\Pi _1}$ is better than partition ${\Pi _2}$ for subset $\mathbb{A}$ \cite{b30}.

Different orders can be adopted to compare relationships between partitions. We adopt Pareto order in this paper to compare the preference relation between two partitions.

\textbf{Definition 3: } For the two partitions ${\Pi _1} = \left\{ {{\cal G}_1^1, \ldots ,{\cal G}_O^1} \right\}$ and ${\Pi _2} = \left\{ {{\cal G}_1^2, \ldots ,{\cal G}_{O'}^2} \right\}$, the utility of  miner $m$ in the partition ${\Pi _1} = \left\{ {{\cal G}_1^1, \ldots ,{\cal G}_O^1} \right\}$ and ${\Pi _2} = \left\{ {{\cal G}_1^2, \ldots ,{\cal G}_{O'}^2} \right\}$ are denoted as ${u_m}\left( {{\Pi _1}} \right)$ and ${u_m}\left( {{\Pi _2}} \right)$, respectively. The partition ${\Pi _1}$ is better than ${\Pi _2}$ by the Pareto order, which is denoted as ${\Pi _1} \triangleright {\Pi _2}$, if and only if
\begin{equation}
{u_m}\left( {{\Pi _1}} \right) \ge {u_m}\left( {{\Pi _2}} \right){\rm{        }}\forall m \subseteq {\Pi _1},{\Pi _2},
\end{equation}
with at least one strict inequality ($>$) for a miner $m$.

A coalition formation algorithm based on the simple rules of merge and split is proposed by using the Pareto order. The coalition formation process usually needs to execute many rounds, and all the coalitions are involved in each round to make sure that the utilities of coalitions increase or remain stable. The rules of merge and split are given as follows \cite{b31}:

\begin{enumerate}
\item Merge Rule: For any set of coalitions $\left\{ {{{\cal G}_1}, \ldots ,{{\cal G}_o}, \ldots ,{{\cal G}_O}} \right\}$, where $\left\{ { \cup _{o = 1}^O{{\cal G}_o}} \right\} \triangleright \left\{ {{{\cal G}_1}, \ldots ,{{\cal G}_o}, \ldots ,{{\cal G}_O}} \right\}$, merge $\left\{ {{{\cal G}_1}, \ldots ,{{\cal G}_o}, \ldots ,{{\cal G}_O}} \right\}$ into $\left\{ { \cup _{o = 1}^L{{\cal G}_o}} \right\}$, which is denoted as $\left\{ {{{\cal G}_1}, \ldots ,{{\cal G}_o}, \ldots ,{{\cal G}_O}} \right\} \to \left\{ { \cup _{o = 1}^O{{\cal G}_o}} \right\}$.
\item Split Rule: For any coalitions $\cup _{o = 1}^O{{\cal G}_o}$, where $\left\{ {{{\cal G}_1}, \ldots ,{{\cal G}_o}, \ldots ,{{\cal G}_O}} \right\} \triangleright \left\{ { \cup _{o = 1}^O{{\cal G}_o}} \right\}$, split $\left\{ { \cup _{o = 1}^O{{\cal G}_o}} \right\}$ into $\left\{ {{{\cal G}_1}, \ldots ,{{\cal G}_o}, \ldots ,{{\cal G}_O}} \right\}$, which is denoted as $\left\{ { \cup _{o = 1}^O{{\cal G}_o}} \right\} \to \left\{ {{{\cal G}_1}, \ldots ,{{\cal G}_o}, \ldots ,{{\cal G}_O}} \right\}$.
\end{enumerate}

The coalitions will merge or split if these actions can yield a preferred collection according to the pareto order. By using the Pareto order, a coalition will split only if the split operation makes at least one miner's utility improved without decreasing other miners' utilities. Similarly, the coalitions will merge only if at least one miner's utility can be increased without hurting other miners' utilities. With the merge-and-split rules, a stable coalition partition can be obtained, as any coalition formation algorithm designed with the merge-and-split rules always converges \cite{b26}. Algorithm 1 is the coalition formation algorithm for miners.
\begin{algorithm}[htp]
\caption{Coalition formation algorithm for miners in the proposed model}
\renewcommand{\algorithmicrequire}{\textbf{Input:}}
\renewcommand{\algorithmicensure}{\textbf{Output:}}
\label{alg1}
\begin{algorithmic}[1]
    \Require Set of miners $\mathbb{M}=\{1,2, \ldots, m, \ldots, M\}$. Workers selected by miners ${W_m} = \left\{ {{{\cal W}_{m,1}},{{\cal W}_{m,2}}, \ldots ,{{\cal W}_{m,\left| {{{\cal W}_m}} \right|}}} \right\}$, $1 \leqslant m \leqslant M$;

    \Ensure The coalition with the highest coalition utility;

    \State Initialization: The partition of miners, where all the miners are disjoint is selected as the initial state. Each miner selects several workers and computes selected workers' compositive reputation value;

    \State Merge mechanism: The Coalition ${{\cal G}_o}$ tries to merge with ${{\cal G}_{o'}}$ based on the merge rule;
    \State Split mechanism: The Coalition ${{\cal G}_o}$ tries to split with ${{\cal G}_{o'}}$ based on the split rule;
    \State Until: Merge and split iteration terminates;
    \State Return: The coalition with the highest coalition utility.
\end{algorithmic}
\end{algorithm}

For the designed coalition formation game $\mathbb{G}$, the grand coalition of all the miners seldom forms due to the communication cost. Besides, the participation of miners that obtain the low reputation values of workers may degrade the coalition utility, which also prevents the grand coalition formation. Next, we use the defection function $\mathbb{D}_{h p}$ to analyze the stability of the final coalition partition.

\textbf{Definition 4: }A partition $\Pi  = \left\{ {{{\cal G}_1},{{\cal G}_2},
\ldots ,{{\cal G}_o}, \ldots ,{{\cal G}_O}} \right\}$ is stable if no
coalition ${{\cal G}_o}$ is incentivized  to change the current partition $\Pi $
by joining another coalition ${{\cal G}_{o'}}$, where ${{\cal G}_o} \cap {{\cal
G}_{o'}} = \phi $, for $o \ne o'$, or trying to split into smaller disjoint coalitions \cite{b25}.

\textbf{Definition 5:} A partition of coalitions $\Pi  = \left\{ {{{\cal G}_1},{{\cal G}_2}, \ldots ,{{\cal G}_o}, \ldots ,{{\cal G}_O}} \right\}$ is $\mathbb{D}_{h p}$-stable if it meets the following two conditions:

1) For $o \in \left\{ {1, \ldots ,O} \right\}$ and each partition $\left\{ {{R_1}, \ldots ,{R_p}} \right\}$ of coalition ${{\cal G}_o}$, we have $\left\{R_{1}, \ldots, R_{p}\right\} \not \triangleright \mathcal{G}_{o}$;

2) For $S \in \left\{ {1, \ldots ,O} \right\}$, we have $\bigcup_{o \in S} \mathcal{G}_{o} \not \triangleright \left\{\mathcal{G}_{o} \mid o \in S\right\}$.
where $\not \triangleright$ is the opposite rule of $ \triangleright $.

\textbf{Theorem 1: } The coalition partition under the proposed scheme is $\mathbb{D}_{h p}$-stable.

{\itshape{Proof}}: We first consider the condition 1. $\Pi  = \left\{ {{{\cal G}_1},{{\cal G}_2}, \ldots ,{{\cal G}_o}, \ldots ,{{\cal G}_O}} \right\}$ is the final partition obtained from the coalition formation algorithm. If for $o \in \left\{ {1, \ldots ,O} \right\}$ and any partition $\left\{ {{R_1}, \ldots, {R_p}} \right\}$ of ${{\cal G}_o}$, there has $\left\{ {{R_1}, \ldots, {R_p}} \right\} \triangleright {{\cal G}_o}$, then the partition ${{\cal G}_o}$ will split, which is in contradiction with the fact that $\Pi$ is a final partition resulted from the merge-and-split iteration. For condition 2, we still consider the same final coalition set $\Pi  = \left\{ {{{\cal G}_1},{{\cal G}_2}, \ldots ,{{\cal G}_o}, \ldots ,{{\cal G}_O}} \right\}$. If for each $S \in \left\{ {1, \ldots ,O} \right\}$, there has $\bigcup_{o \in S} \mathcal{G}_{o} \not \triangleright \left\{\mathcal{G}_{o} \mid o \in S\right\}$, then the partition $\Pi$ can be modified through the merge rule, which is also in contradiction with the fact that $\Pi$ is a final stable partition. Thus, the coalition partition under the proposed scheme is $\mathbb{D}_{h p}$-stable.

\section{STACKELBERG GAME-BASED INCENTIVE MECHANISM FOR VEHICULAR METAVERSE}
After the worker selection phase, MSP ${P_i}$ obtains the information of workers in the selected coalition ${{\cal G}_o}$. MSP ${P_i}$ selects $N$ workers from the coalition ${{\cal G}_o}$ based on the workers' reputation opinions. The set of selected workers is denoted as $\mathbb{W}_{\text {sel }}=\{1,2, \ldots, w, \ldots, N\}$, and ${\mathbb{W}_{{\text{sel}}}} \subseteq {{\cal G}_o}$. In the CDC rendering task execution phase shown in Fig. 2, MSP ${P_i}$
divides the matrix ${A_i}$ into $K$ equal-sized submatrices
${\mathbb{R}^{\frac{{(3\times2\times{n_p})}}{K} \times {n_p}}}$, and then MSP ${P_i}$ has $N$ encoded submatrices with the same size $\frac{{(3\times2\times{n_p})}}{K} \times {n_p}$ based on the $\left( {N,K} \right)$ MDS code. Each submatrix is allocated to a worker. The distributed computing resource interactions between MSP ${P_i}$ and workers is modeled as a single-leader multi-followers Stackelberg game. In the leader
game, the MSP selects an optimal computation reward to motivate workers to execute CDC rendering tasks in the vehicular metaverse. In the follower game, the workers try to obtain higher profit by adjusting their computing speed. The Stackelberg game model can be extended to multiple MSPs with two approaches. The first one is that the MSPs make decisions on incentive indenpendently and they do not affect each other as the MSPs may use different applications. The second one is that MSPs can compete with each other and this forms a Nash game which can be investigated in the future work.

\subsection{Profit Function of Workers}
We consider that the CDC rendering task execution mainly contains computing the task and transmitting the computation result to  MSP ${P_i}$. In the following, the profit function of workers is studied.

In order to incentivize workers to join the CDC rendering task actively,
the MSP ${P_i}$ gives the reward to workers that contribute to vehicular metaverse services. The reward contains two kinds, i.e., base reward ${{\cal R}_{base}}$ and competition reward ${{\cal R}_{com}}$. The workers who participate in the CDC rendering task can
receive the reward ${{\cal R}_{base}}$. The workers whose average task execution time
is no more than the average task execution time of the $K$-th worker can receive the reward ${{\cal R}_{com}}$. Similar to the analysis in \cite{c3}, we assume that the task execution time of workers follows a uniform distribution, where $\frac{{{T_w}}}{{{T_{\max }}}} \in \left( {0,1} \right)$, and ${{T_{\max }}}$ is the maximum value of the task execution time. The normalized task execution time of workers are ranked and represented by its order statistics, which are expressed as ${T_{1,N}},{T_{2,N}}, \ldots ,{T_{N,N}}$. ${T_{K,N}}$ is the $K$-th highest execution time among $N$ workers. The cumulative distribution function of the normalized task execution time is $F\left( T \right) = T$. The probability density function of the normalized task execution time is $f\left( T \right) = 1$. Based on the order statistics, the probability density function of ${T_{k,N}}$ is given as
\begin{equation}
{f_{\left( k \right)}}\left( T \right) = Nf\left( T \right)\left( {\begin{array}{*{20}{c}}
  {N - 1} \\
  {k - 1}
\end{array}} \right)F{\left( T \right)^{k - 1}}{\left( {1 - F\left( T \right)} \right)^{N - k}},
\end{equation}
which is also a beta distribution ${Beta}\left( {k,N - k + 1} \right)$. Hence, the expectation of ${T_{k,N}}$ is given as
\begin{equation}
\mathbb{E}\left( {{T_{K,N}}} \right) = \frac{K}{{N + 1}}.
\end{equation}
The workers adopt the dynamic voltage scaling technology that allows
workers to adaptively adjust and control the computing speed \cite{b32}. Each selected worker tries to obtain a higher profit.
The profit function of the worker $w$ is expressed as
\begin{equation}
{u_{w}} = {\mathcal{R}_{base}} + {\mathcal{P}_w}{\mathcal{R}_{com}} - {\varepsilon}{\mu_w}\mathbb{E}\left( {T_w^{cmp}} \right) - {\zeta}\mathbb{E}\left( {T^{com - u}} \right),
\end{equation}
where ${\varepsilon}$ is the computing cost of the worker per CPU circle, and ${\zeta}$ is the communication cost of the worker per unit of communication time. ${{\cal P}_w}$ is the probability of worker $w$ getting the
reward from MSP ${P_i}$. ${{\cal P}_w}$ is expressed as
\begin{equation}
\begin{gathered}
  {\mathcal{P}_w} = 1 - {e^{ - {\mu _w}\left( {\frac{{\mathbb{E}\left( {{T_{K,N}}} \right) \times {T_{\max }} - \mathbb{E}\left( {T_{K,N}^{com - u}} \right)}}{{{l}}} - {a_w}} \right)}}, \hfill \\
  {\text{   {\kern 1pt}{\kern 1pt}{\kern 1pt}{\kern 1pt}{\kern 1pt}{\kern 1pt}{\kern 1pt}{\kern 1pt}{\kern 1pt}{\kern 1pt}    = 1}}-{e^{ - {\mu _w}{A_w}}}, \hfill \\
\end{gathered}
\end{equation}
where ${\mathbb{E}\left( {T_{K,N}^{com - u}} \right)}$ is the average communication delay of the $K$-th worker, and ${A_w} = \frac{{\mathbb{E}\left( {{T_{k,N}}} \right) \times {T_{\max }} - \mathbb{E}\left( {T_{k,N}^{com - u}} \right)}}{{{l}}} - {a_w}$.
Here, worker $w$ selects the optimal computing speed ${\mu _w}$ that maximizes ${u_w}$. 
\subsection{Utility Function of Metaverse Service Provider}
To motivate workers to contribute more to the CDC rendering tasks, MSP ${P_i}$
should adjust the reward ${{\cal R}_{com}}$ to maximize its utility. The selected
workers contribute their computing resources to MSP ${P_i}$. The utility that MSP ${P_i}$ can gain
depends on the computing resource that workers contribute, the reputation values of workers, and the reward paid to the workers. The utility function of MSP ${P_i}$ is expressed as
\begin{equation}
{u_{{P_i}}} = \nu \sum\nolimits_{w = 1}^N {f\left( {{\mu _w}} \right)h\left( {T_{i \to w}^{com}} \right)}  - N{\mathcal{R}_{base}} - \sum\nolimits_{w = 1}^N {{\mathcal{P}_w}{\mathcal{R}_{com}}},
\end{equation}
where $\nu $ is a weight parameter, and $f\left( {{\mu _w}} \right) = \log \left( {1 + {\mu _w}} \right)$, which is the utility of MSP ${P_i}$ gained from the workers' computation contribution. The log of $f\left(  \cdot  \right)$ reflects MSP ${P_i}$'s diminishing
return on the computation speed of each selected worker\cite{b33,b34,b35}. $h\left(  \cdot  \right)$ is the reputation utility of workers and is expressed as \cite{b36}
\begin{equation}
h\left( {T_{i \to w}^{com}} \right) = \alpha  + \left( {1 - \alpha } \right)\log \left( {1 + \frac{{\left( {e - 1} \right)\left( {T_{i \to w}^{com} - T_{th}^{com}} \right)}}{{T_{\max }^{com} - T_{th}^{com}}}} \right).
\end{equation}
where $\alpha$ is the default reputation utility for the worker with $T_{i \to w}^{com} = T_{th}^{com}$,  $T_{th}^{com}$ is the reputation threshold required by the MSP. $T_{\max }^{com}$ is the maximum reputation value. 
\subsection{Stackelberg Game-Based Incentive Mechanism}
The interactions among the MSP and workers is formulated as a single-leader multi-followers Stackelberg game. In the leader game, the MSP sets the optimal computation reward to motivate workers to execute CDC rendering tasks. In the follower game, the coalition of workers adjust the computation speed to maximize their profits. The strategy optimization problems for the MSP and workers are formulated as follows.
\begin{enumerate}
\item Workers' computing speed strategies in Stage II: In Stage II, based on the reward strategy of the MSP, worker $w$ $({w \in {\mathbb W_{{\text{sel}}}}})$ determines its computation speed ${\mu _w}$ to maximize the profit that is given as
\begin{equation}
\begin{array}{l}
{u_w}\left( {{\mu _w};{{\cal R}_{com}}} \right) = {{\cal R}_{base}} + {{\cal P}_w}\left( {{\mu _w}} \right){{\cal R}_{com}}\\{\kern 72pt}
- \varepsilon {\mu _w}\mathbb{E}\left( {T_w^{cmp}} \right) - {\zeta}\mathbb{E} \left( {{T^{com\_u}}} \right)
\end{array}.
\end{equation}
The set of workers' computing speed is $\boldsymbol{\mu}  = \left\{ {{\mu _1}, \ldots ,{\mu _w}, \ldots {\mu _N}} \right\}$, which are used to derive the computation delay of workers based on (7). The worker subgame problem is expressed as follows.

\textbf{Problem 1} (Worker $w$ Subgame):
\begin{equation}
\begin{aligned}
&\underset{\mu_{w}}{\operatorname{maximize}} {\kern 2pt}{\kern 2pt}{\kern 2pt}{\kern 2pt} {\kern 2pt} u_{w,{w \in {\mathbb{W}_{{\text{sel}}}}}}\left(\mu_{w} ; \mathcal{R}_{\mathrm{com}}\right) \\
&\text { subject to } \quad \underline{\mu} \leq \mu_{w} \leq \bar{\mu},
\end{aligned}
\end{equation}
where $\underline{\mu}$ is the minimum computation speed, and $\bar{\mu}$ is the maximum computation speed.
\item MSP's reward strategy in Stage I: In Stage I, according to workers' computation speed strategies $\boldsymbol{\mu}$, the MSP determines the reward strategy to maximize its utility that is expressed as
\begin{equation}
\begin{gathered}
  {u_{{P_i}}}\left( {{\mathcal{R}_{com}};\boldsymbol{\mu} } \right) = \nu \sum\nolimits_{w = 1}^N {f\left( {{\mu _w}} \right)h\left( {T_{i \to w}^{com}} \right)}  - N{\mathcal{R}_{base}} \hfill \\
   {\kern 70pt}- \sum\nolimits_{w = 1}^N{{\mathcal{P}_w}{\mathcal{R}_{com}}}  \hfill \\
\end{gathered}.
\end{equation}
The MSP subgame problem is expressed as follows.

\textbf{Problem 2} (MSP ${P_i}$ Subgame)
\begin{equation}
\begin{aligned}
&\underset{\quad R_{\text {com }}}{\text { maximize }} u_{P_{i}}\left(\mathcal{R}_{\text {com }} ; \boldsymbol{\mu}\right) \\
&\text { subject to } \quad \underline{\mathcal{R}}_{\text {com }} \leq \mathcal{R}_{\text {com }} \leq \overline{\mathcal{R}}_{\text {com }},
\end{aligned}
\end{equation}
where $\underline{\mathcal{R}}_{\text {com }} $ is the minimum computation reward, $\overline{\mathcal{R}}_{\text {com }}$ is the maximum computation reward.
\end{enumerate}
The Stackelberg game is formulated by combining Problem 1 and Problem 2, and the goal of the Stackelberg game is to find the Nash equilibrium solution.
\subsection{Game Equilibrium Analysis}
The Stackelberg equilibrium makes sure that the utility of the MSP is maximized considering that the workers contribute their computation resource to vehicular metaverse services based on the best response. This means that the computing speed strategy of each worker
maximize its profit given the strategies of the other workers and the computing reward given by the MSP. The Stackelberg equilibrium is expressed as follows.

\textbf{Definition 6:} We denote ${\boldsymbol{\mu}^*}$ and ${\cal R}_{com}^*$ as the optimal computation speed of all the selected workers and optimal computation reward given by the MSP, respectively. Then, the strategy $\left( {{\boldsymbol{\mu} ^*},{\cal R}_{com}^*} \right)$ is the Stackelberg equilibrium if we have
\begin{equation}
{u_{{P_i}}}\left( {{\cal R}_{com}^*;{\boldsymbol{\mu}^*}} \right) \ge {u_{{P_i}}}\left( {{{\cal R}_{com}};{\boldsymbol{\mu} ^*}} \right),
\end{equation}
\begin{equation}
{u_w}\left( {\mu _w^*;\boldsymbol{\mu} _{ - w}^*,{\cal R}_{com}^*} \right) \ge {u_w}\left( {{\mu _w};\boldsymbol{\mu} _{ - w}^*,{\cal R}_{com}^*} \right),  \forall w \in \mathbb{W}_\text{sel}.
\end{equation}

The backward induction is adopted to analyze the Stackelberg game.

\emph{1) Workers' optimal strategies as equilibrium in Stage II}

 Based on the reward strategy ${{\cal R}_{com}}$ given by the MSP, the workers determine the optimal computation speed strategy for profit maximization in Stage II.

\textbf{Theorem 2: } The sub-game perfect equilibrium in the workers' subgame is unique.

{\itshape{Proof}}: We give the first-order and second-order derivatives of the profit function of workers ${u_w}\left(  \cdot  \right)$ with respect to worker's strategy ${\mu _w}$. The first-order derivative of ${u_w}\left(  \cdot  \right)$ is shown as
\begin{equation}
\begin{array}{l}
\frac{{\partial {u_w}}}{{\partial {\mu _w}}} =   {{{\cal R}_{com}}{A_w}{e^{ - {\mu _w}{A_w}}} - {\varepsilon }{l}{a_w}{e^{{a_w}{\mu _w}}}}.
\end{array}
\end{equation}
 The second derivative of ${u_w}\left(  \cdot  \right)$ is shown as
\begin{equation}
\begin{array}{l}
\frac{{{\partial ^2}{u_w}}}{{\partial {\mu _w}^2}} =  - \left( {{A^2_w}{e^{ - {\mu _w}{A_w}}}{{\cal R}_{com}} + {\varepsilon}{l}{a^2_w}{e^{{a_w}{\mu _w}}}} \right) < 0.
\end{array}
\end{equation}

As the second-order derivative of ${u_w}\left(  \cdot  \right)$ is negative, so the profit function ${u_w}\left(  \cdot  \right)$ is strictly concave with respect to worker's computing speed strategy ${\mu _w}$. Besides, based on the first-order derivative condition, there is
\begin{equation}
\begin{array}{l}
\frac{{\partial {u_w}}}{{\partial {\mu _w}}} =  {{{\cal R}_{com}}{A_w}{e^{ - {\mu _w}{A_w}}} - {\varepsilon }{l}{a_w}{e^{{a_w}{\mu _w}}}} \ = 0.
\end{array}
\end{equation}
Then,  the best response function of the worker $w$, i.e., $\mu _w^*$ , is shown as
\begin{equation}
\begin{array}{l}
\mu _w^* = \frac{1}{{{a_w} + {A_w}}}\log \left[ {\frac{{ {{\cal R}_{com}}{A_w} }}{{ {\varepsilon }{l}{a_w}}}} \right].
\end{array}
\end{equation}
We denote ${E_w} = \frac{1}{{{a_w} + {A_w}}}$ and ${F_w} = \frac{{ {A_w}}}{{ {\varepsilon }{l}{a_w}}}$. Thus, the sub-game perfect equilibrium of the workers' subgame is unique \cite{c4}.
\begin{figure*}[t!]
\begin{equation}
{u_{{P_i}}}\left( {{\mathcal{R}_{com}};\mu } \right) = U\left( {\mu _1^*, \ldots ,\mu _w^*, \ldots ,\mu _N^*} \right) - N{\mathcal{R}_{base}} - \sum\nolimits_{w = 1}^N {\left[ {1 - {{\left( {{F_w}{\mathcal{R}_{com}} } \right)}^{ - {E_w}{A_w}}}} \right]{\mathcal{R}_{com}}}.
\end{equation}
\end{figure*}
\begin{figure*}[t!]
\begin{equation}
\frac{{\partial {u_{{P_i}}}}}{{\partial {{\cal R}_{com}}}} = \frac{{\partial U\left( {\mu _1^*, \cdots ,\mu _w^*, \cdots ,\mu _N^*} \right)}}{{\partial {{\cal R}_{com}}}} - \sum\nolimits_{w = 1}^N {\left[ {1 + \left( {{E_w}{A_w} - 1} \right){{\left( {{F_w}{{\cal R}_{com}}} \right)}^{ - {E_w}{A_w}}}} \right]} .
\end{equation}
\end{figure*}
\begin{figure*}[t!]
\begin{equation}
\frac{{{\partial ^2}{u_{{P_i}}}}}{{\partial {\cal R}_{com}^2}} = \frac{{{\partial ^2}U\left( {\mu _1^*, \cdots ,\mu _w^*, \cdots ,\mu _N^*} \right)}}{{\partial {\cal R}_{com}^2}} - \sum\nolimits_{w = 1}^N {\left[ { - {E_w}{A_w}{F_w}\left( {{E_w}{A_w} - 1} \right){{\left( {{F_w}{{\cal R}_{com}}} \right)}^{ - {E_w}{A_w} - 1}}} \right]}.
\end{equation}
\end{figure*}

\emph{2) MSP's optimal reward strategy in Stage I}

Based on the optimal computing speed strategies of workers in Stage II, the MSP acts as the leader to optimize its utility in Stage I.

\textbf{Theorem 3: }  The uniqueness of the proposed Stackelberg game's equilibrium can be guaranteed.

{\itshape{Proof}}:  The utility function of the MSP can be transformed into (36), and $U\left( {\mu _1^*,\mu _2^*, \ldots ,\mu _w^*, \ldots ,\mu _N^*} \right) = \nu \sum\nolimits_{w = 1}^N {f\left( {\mu _w^*} \right)h\left( {T_{i \to w}^{com}} \right)}$. The first-order derivative of the MSP's utility function is shown in (37). The second-order derivative of the MSP's utility function is shown in (38). As $\frac{{{\partial ^2}U\left(  \cdot  \right)}}{{\partial {\cal R}_{com}^2}} < 0$, when ${E_w}{A_w} < 1$, there is $\frac{{{\partial ^2}{u_{{P_i}}}}}{{\partial {\cal R}_{com}^2}} < 0$. Because ${E_w}{A_w} = \frac{{{A_w}}}{{{a_w} + {A_w}}} = \frac{1}{{\frac{{{a_w}}}{{{A_w}}} + 1}} < 1$, the second-order derivative of the MSP's utility function always satisfies $\frac{{{\partial ^2}{u_{{P_i}}}}}{{\partial {\cal R}_{com}^2}} < 0$, which indicates that ${u_{{P_i}}}$ is a concave function. Thus, the MSP has a unique optimal solution that can be efficiently obtained by bisection method \cite{b37}. Based on the optimal strategy of the MSP, the workers' optimal strategies can be obtained. Then, the Stackelberg equilibrium can be obtained in the proposed model. The MSP can achieve optimal utility and workers can obtain optimal profit, and neither of them would alter their strategies to gain higher benefits.

We design Algorithm 2 to obtain the unique Stackelberg equilibrium for the proposed game. In algorithm 2, the MSP first searches the optimal reward strategy, and the complexity is $O\left( {{{\log }_2}\left( {{{\overline {\cal R}}_{com}} - {\underline {\cal R}_{com}}} \right)} \right)$. Then, each worker needs to decide its optimal computing speed strategy, the complexity is $O\left( N \right)$. So the algorithm complexity is $O\left( {N + {{\log }_2}\left( {{{\overline {\cal R}}_{com}} - {{\underline{\cal R}}_{com}}} \right)} \right)$.

\begin{algorithm}[htp]
	\caption{Bisection-based algorithm to find Stackelberg equilibrium}
	\renewcommand{\algorithmicrequire}{\textbf{Input:}}
	\renewcommand{\algorithmicensure}{\textbf{Output:}}
	\label{alg1}
	\begin{algorithmic}[1]
		\Require Minimum computation reward $\underline{\mathcal{R}}_{\text {com }} $, maximum computation reward $\overline{\mathcal{R}}_{\text {com }}$, error $\Xi $ ;
		
		\Ensure Optimal reward strategy of the MSP ${\cal R}_{com}^*$ and optimal computing speed of workers ${\boldsymbol{\mu}^*}$;

		\State MSP find the optimal reward strategy based on the following steps.
		
		\While {${{\partial {u_{{P_i}}}} \mathord{\left/
					{\vphantom {{\partial {u_{{P_i}}}} {\partial {{\cal R}_{com}}}}} \right.
					\kern-\nulldelimiterspace} {\partial {{\underline{\cal R}}_{com}}}} \times {{\partial {u_{{P_i}}}} \mathord{\left/
					{\vphantom {{\partial {u_{{P_i}}}} {\partial {{\bar {\cal R}}_{com}}}}} \right.
					\kern-\nulldelimiterspace} {\partial {{\overline {\cal R}}_{com}}}} < 0$
			}
		\State ${{\cal R}_{com}} = 0.5 \times \left( {\underline{\mathcal{R}}_{\text {com }} + {{\overline {\cal R}}_{com}}} \right)$;
		\If {${{\partial {u_{{P_i}}}} \mathord{\left/
					{\vphantom {{\partial {u_{{P_i}}}} {\partial {{\cal R}_{com}}}}} \right.
					\kern-\nulldelimiterspace} {\partial {{\underline{\cal R}}_{com}}}} \times {{\partial {u_{{P_i}}}} \mathord{\left/
					{\vphantom {{\partial {u_{{P_i}}}} {\partial {{\bar {\cal R}}_{com}}}}} \right.
					\kern-\nulldelimiterspace} {\partial {{ {\cal R}}_{com}}}} < 0$}
		\State ${\overline {\cal R}_{com}} = {{\cal R}_{com}}$;
		\Else
		\State ${{{{\underline {\cal R}}_{com}}}} = {{\cal R}_{com}}$;
		\EndIf
		\If {$\left| {{{\overline {\cal R}}_{com}} - {{\underline {\cal R}}_{com}}} \right| < \Xi $}
		\State break;
		\EndIf
		\EndWhile
		\State ${\cal R}_{com}^* = 0.5 \times \left( {{{\overline {\cal R}}_{com}} + {{\underline {\cal R}}_{com}}} \right)$;
		\State Each worker $w$ decides its computing speed strategy $\mu _w$ based on (35);
		\State Return: Optimal reward strategy of the MSP ${\cal R}_{com}^*$ and optimal computing speed of workers ${\boldsymbol{\mu}^*}$;
\end{algorithmic}
\end{algorithm}

\begin{table}[htpb] 
\caption{Parameter settings}

\begin{tabular}{c|c}
\toprule
Parameter	                    &  Value       \\\hline
Number of workers $W$ & $100$\\
Number of miners $M$ &  $20$\\
Weight of positive and negative events ${\sigma _1}$, ${\sigma _2}$   &  $0.6$, $0.4$            \\
Packet size ${s^u}$ & $400B$\\
Bandwidth assigned to workers $B$ & 10MHz  \\
pathloss exponent       &  $3$  \\
Log-normal shadowing standard deviation                &   3dB   \\
Variance of the Gaussian noise ${{{\tilde \sigma }^2}}$ & -30dBm \\
Maximum transmitting power of workers $p$ & 5w\\
Feedback period of mobile devices $T_c$  & 0.1ms \\
Threshold of reputation value $T_{th}^{com}$ &  $ 0.6$ \\
Workers' computation cost per CPU circle $\varepsilon $ & $0.1$\\
workers' communication cost per unit time $\zeta$ & $10$ \\
Default parameters $\nu$ &    10\\
 Maximum relative speed between MSP and workers & $15m/s$\\

\bottomrule
\end{tabular}
\end{table}
\section{PERFORMANCE ANALYSIS}
In this section, we give the numerical results of the coalition game-based worker selection and stackelberg game-based incentive mechanism for reliable CDC in the vehicular metaverse. The parameter values are given in Table I \cite{b15} \cite{vehicle_mobility_1}.

\subsection{Numerical Results for Reliable worker selection}
First, the reputation calculation scheme is simulated and analyzed. Then, the coalition formation game is analyzed. 
For the reputation calculation scheme, we consider an unreliable worker, which performs well to all MSPs to increase its reputation value to 0.8 at first, and keeps such reputation value for a certain period of time. Then, the unreliable worker keeps performing well to several specific MSPs, but misbehaves to other MSPs with the probability of 90\%. The proposed blockchain-enabled reputation scheme is compared with the reputation scheme without blockchain, and the reputation scheme without blockchain and recommended opinions. For the reputation scheme without blockchain, such as the reputation scheme in \cite{reputation_centre}, workers' reputation values are all stored in the centralized platform in which unreliable workers' negative interaction behaviors are manipulated into positive interaction behaviors with
the probability of 25\%. For the reputation scheme without blockchain and recommended opinions, workers' reputation values only depend on MSPs' local reputation opinions.
\begin{figure}[t]
\begin{center}
\includegraphics[width=0.4\textwidth]{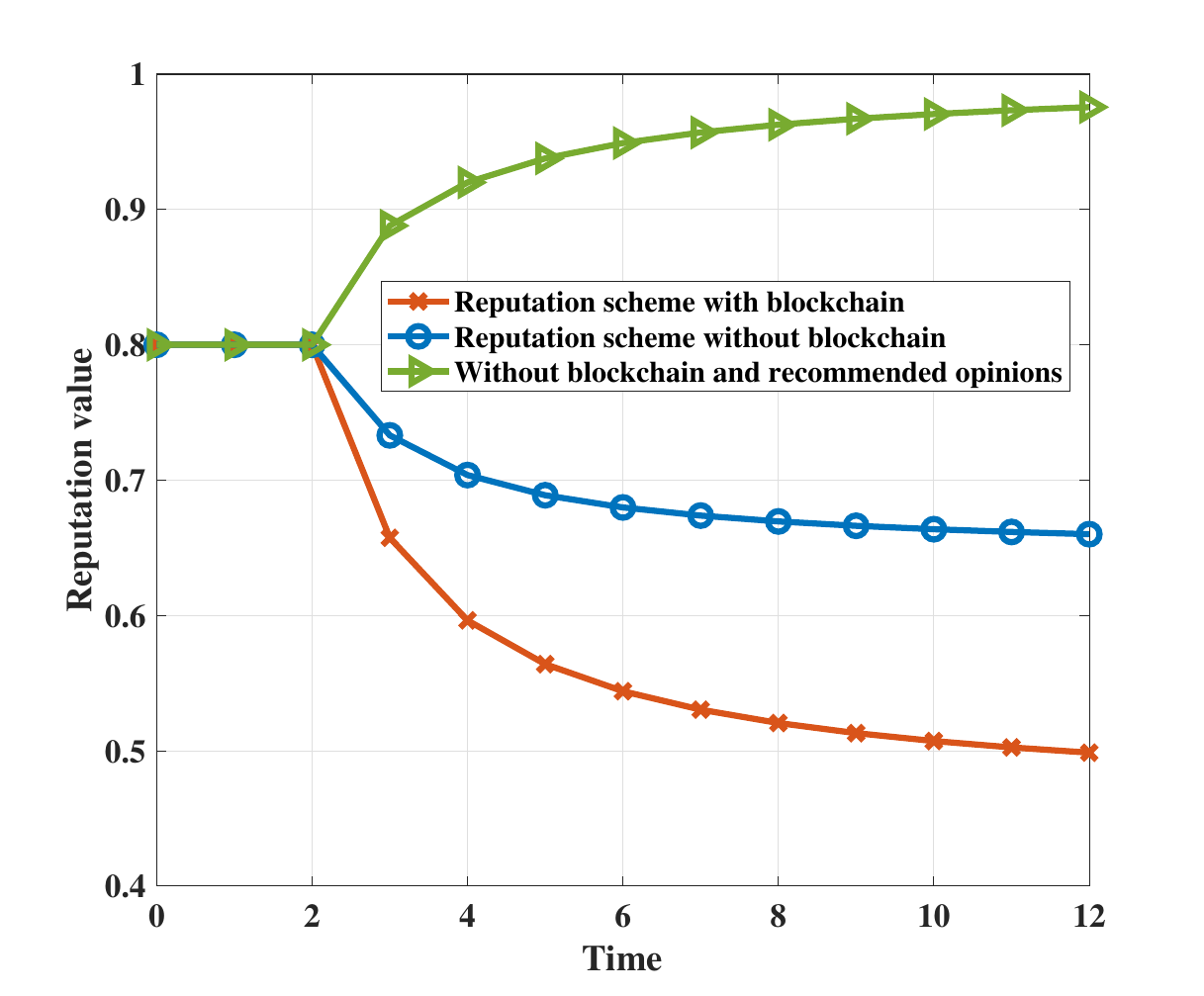}
\end{center}
\caption{Reputation values variation of an unreliable worker.}
\label{Fig4}
\end{figure}
\begin{figure}[t]
\begin{center}
\includegraphics[width=0.4\textwidth]{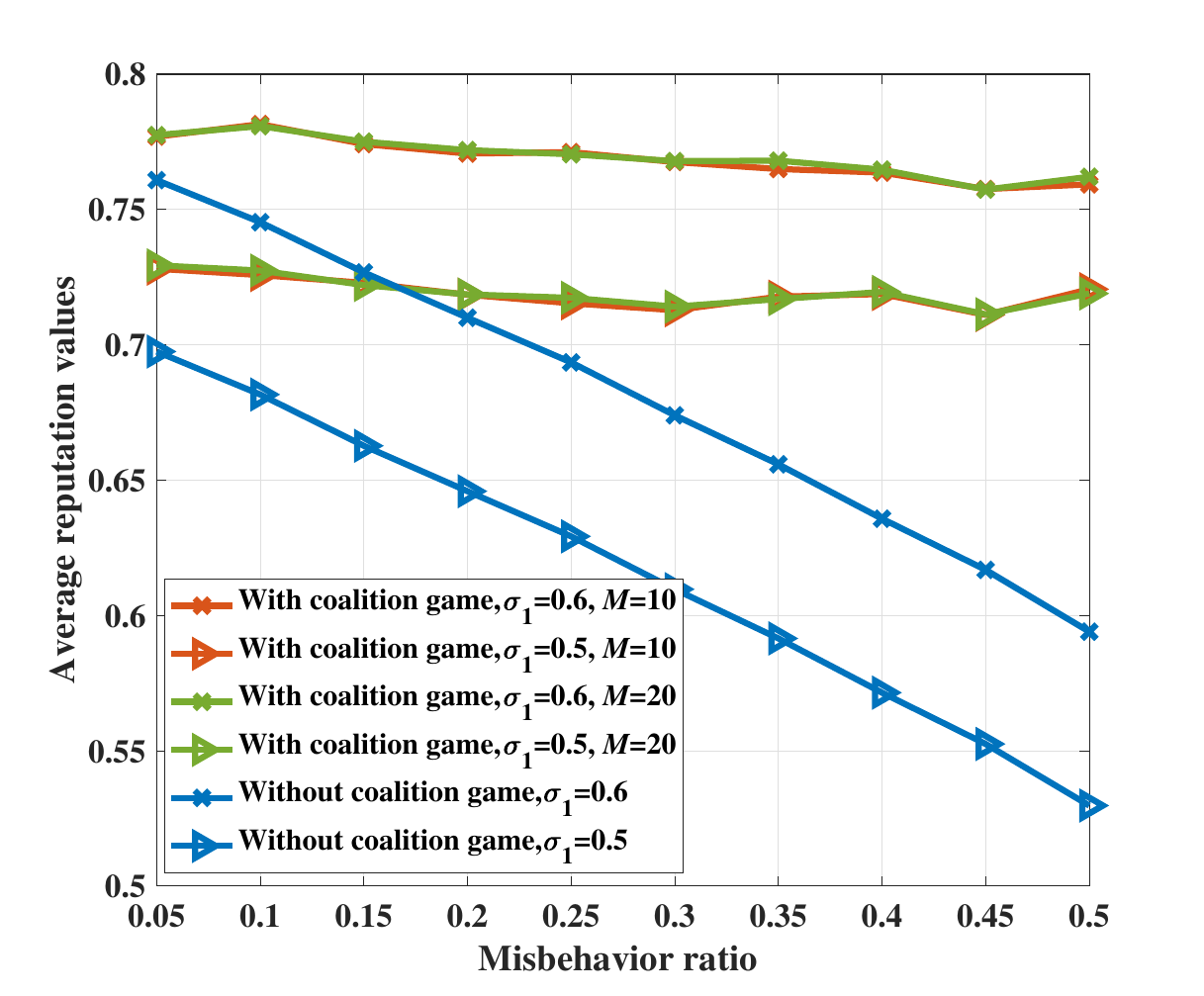}
\end{center}
\caption{Average reputation values as function of misbehavior ratio.}
\label{Fig5}
\end{figure}
\begin{figure}[t]
	\begin{center}
		\includegraphics[width=0.4\textwidth]{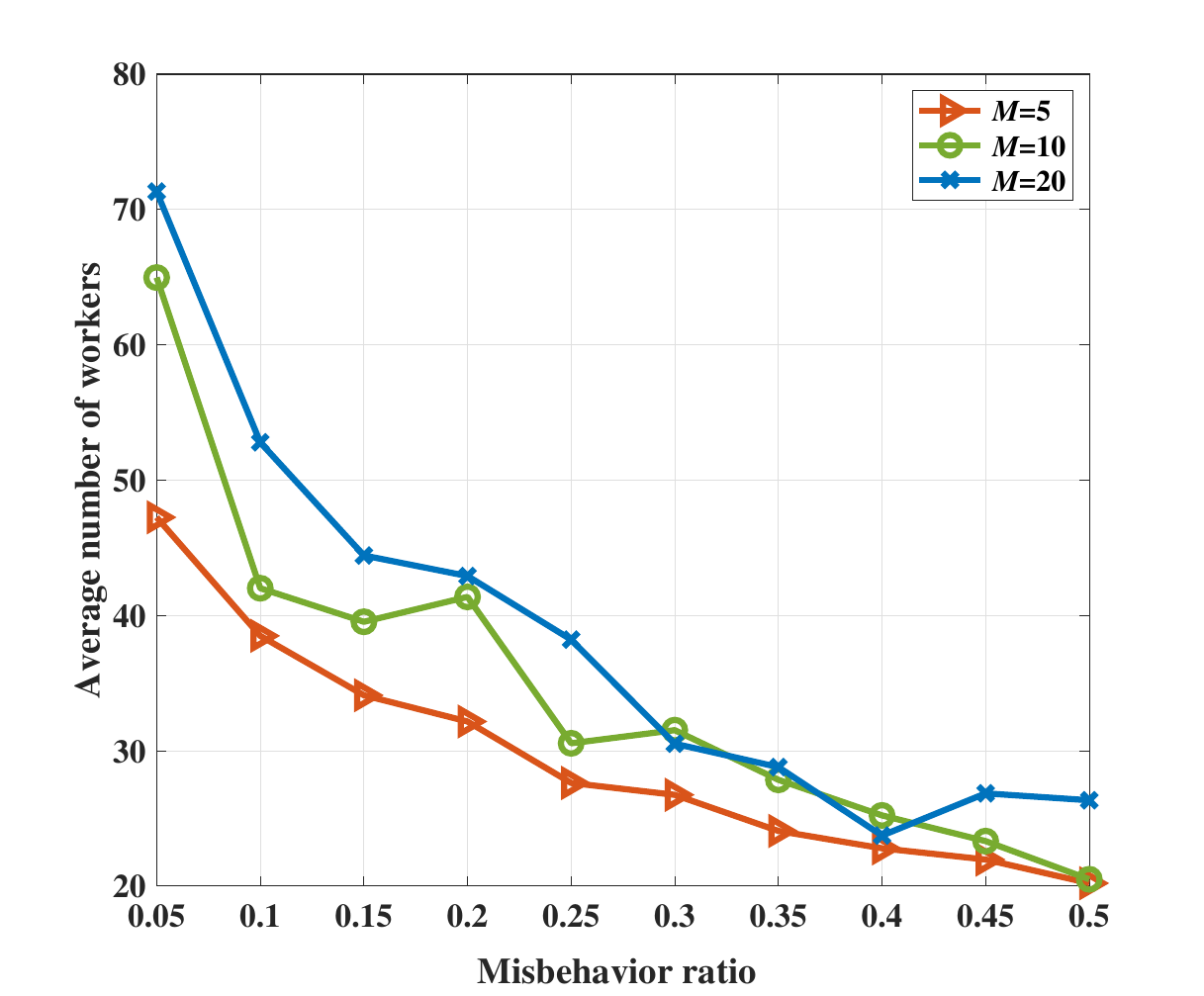}
	\end{center}
	\caption{Number of selected workers as function of misbehavior ratio.}
	\label{Fig6}
\end{figure}
Figure 4 shows the reputation values variation of an unreliable worker over time. As it can be seen from Fig. 4, when the unreliable worker begins to misbehave, the reputation value significantly decreases with the proposed blockchain-enabled reputation scheme. For the reputation scheme without blockchain, the reputation value decreases more slowly than that of the reputation scheme with blockchain. As the centralized platform manipulates the unreliable workers' negative interactions into positive interactions, which increases the unreliable worker's reputation value. For the reputation scheme without blockchain and recommended opinions, the reputation value of the unreliable worker still increases, as the MSPs that are served positively by the unreliable worker compute the reputation value only based on local reputation opinions.

Next, we analyze the coalition formation game. We set the number of positive interaction events as $0 \sim 120$, and the number of negative interaction events as $0\sim 40$. Then the compositive reputation values of workers are calculated by miners based on subjective logical model. Figure 5 shows the selected workers' average reputation value as a function of misbehavior ratio.
Misbehavior ratio is the percentage of workers that are misbehaving to the total number of workers. From Fig. 5, compared with the scheme without coalition game, the selected workers' average reputation value of the proposed scheme decreases slightly with the growth of misbehavior ratio. The misbehavior ratio has little effect on the average reputation value of workers in the selected worker coalition, as the coalition game-based method helps to exclude the workers with low reputation values.
When the weight of positive event ${\sigma _1}$ is fixed, the number of miners joining in the reputation calculation does not affect the average reputation values of selected workers.
When the misbehavior ratio is fixed, the selected workers' average reputation value increases with the increase of ${\sigma _1}$. 
Figure 6 shows the number of selected workers as function of misbehavior ratio. From Fig. 6, when the number of miners is fixed, the number of selected workers decreases with the rise of misbehavior ratio. When the misbehavior ratio is fixed, the number of selected workers increases with the growth of the number of miners. As more miners can calculate more workers' reputation values, which is benefical to the reliable worker selection.

\subsection{Numerical Analysis for CDC Incentive Scheme}
\begin{figure}[t]
\begin{center}
\includegraphics[width=0.4\textwidth]{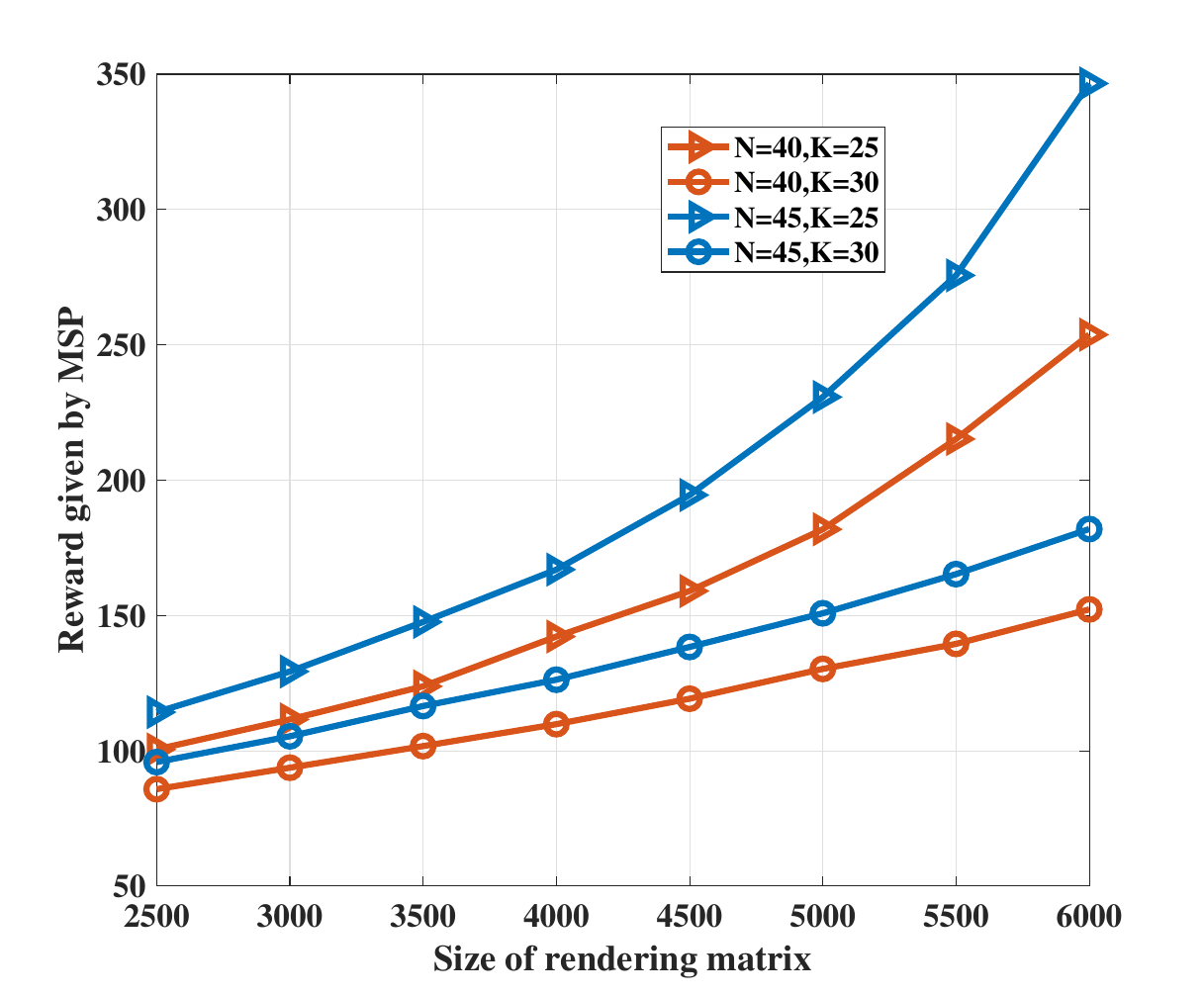}
\end{center}
\caption{Reward strategy of MSP as function of size of rendering matrix.}
\label{Fig7}
\end{figure}
\begin{figure}[t]
\begin{center}
\includegraphics[width=0.4\textwidth]{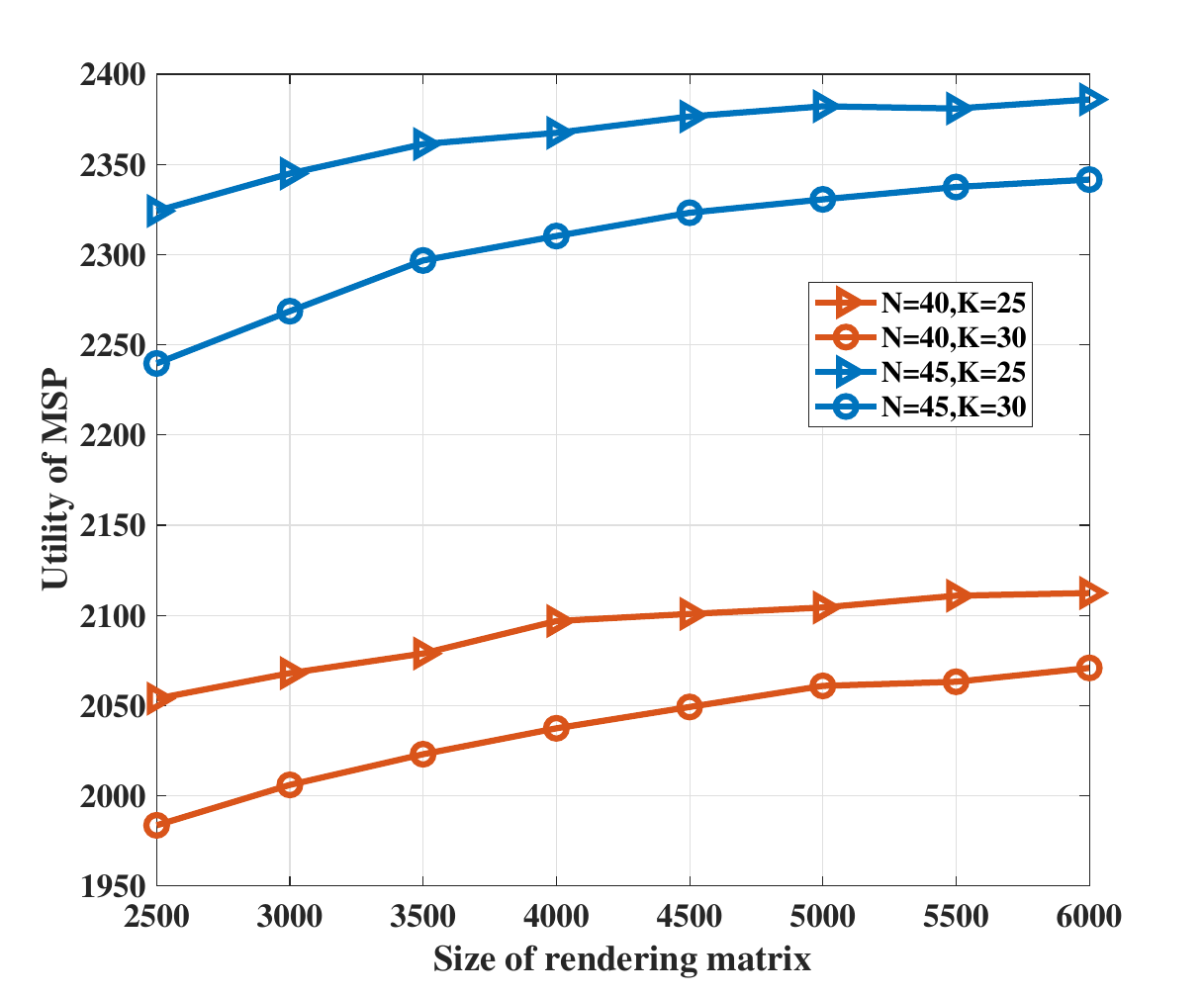}
\end{center}
\caption{Utility of MSP as function of size of rendering matrix.}
\label{Fig8}
\end{figure}
\begin{figure}[t]
\begin{center}
\includegraphics[width=0.4\textwidth]{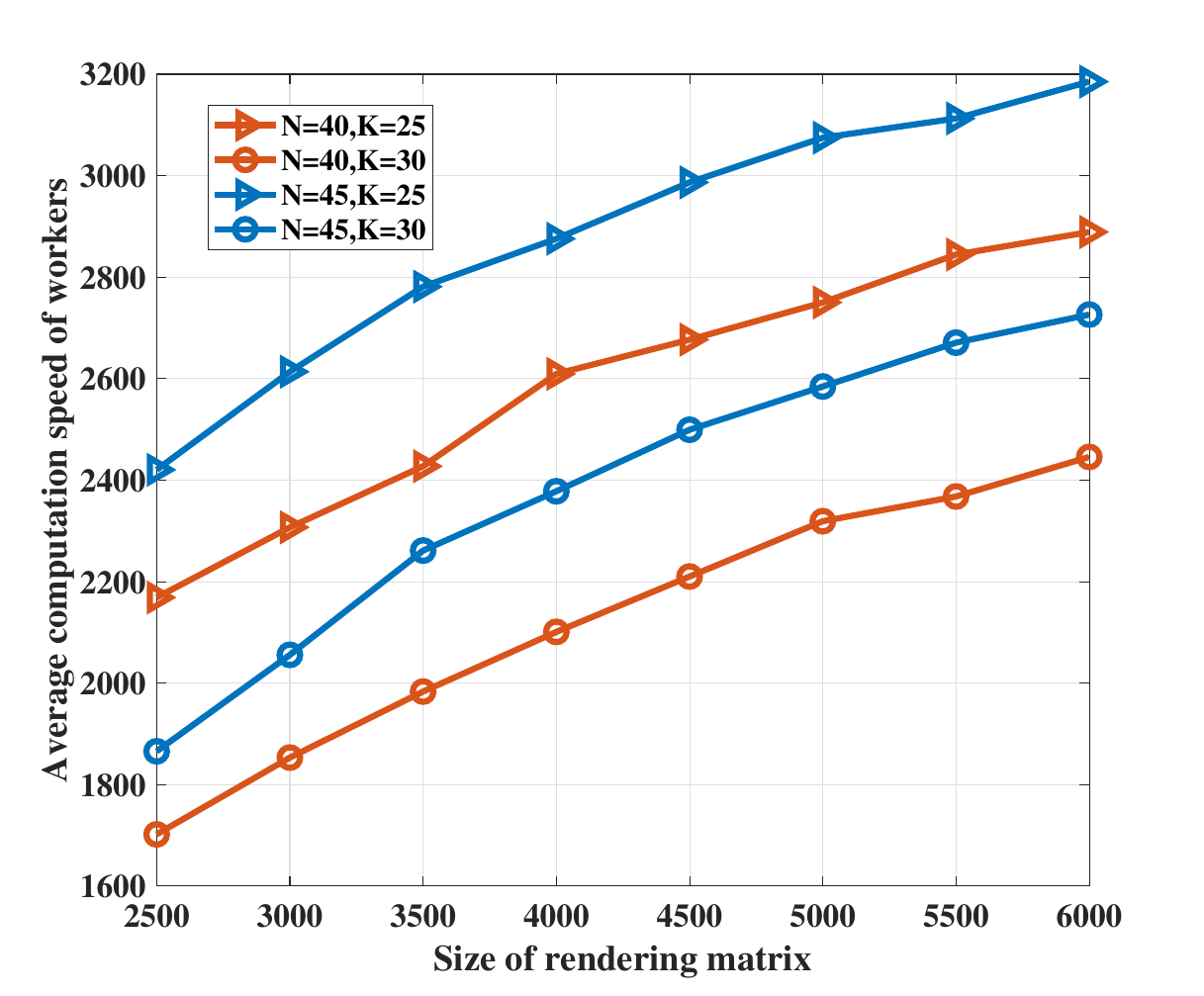}
\end{center}
\caption{Average computation speed of workers as function of size of rendering matrix.}
\label{Fig9}
\end{figure}
\begin{figure}[t]
\begin{center}
\includegraphics[width=0.4\textwidth]{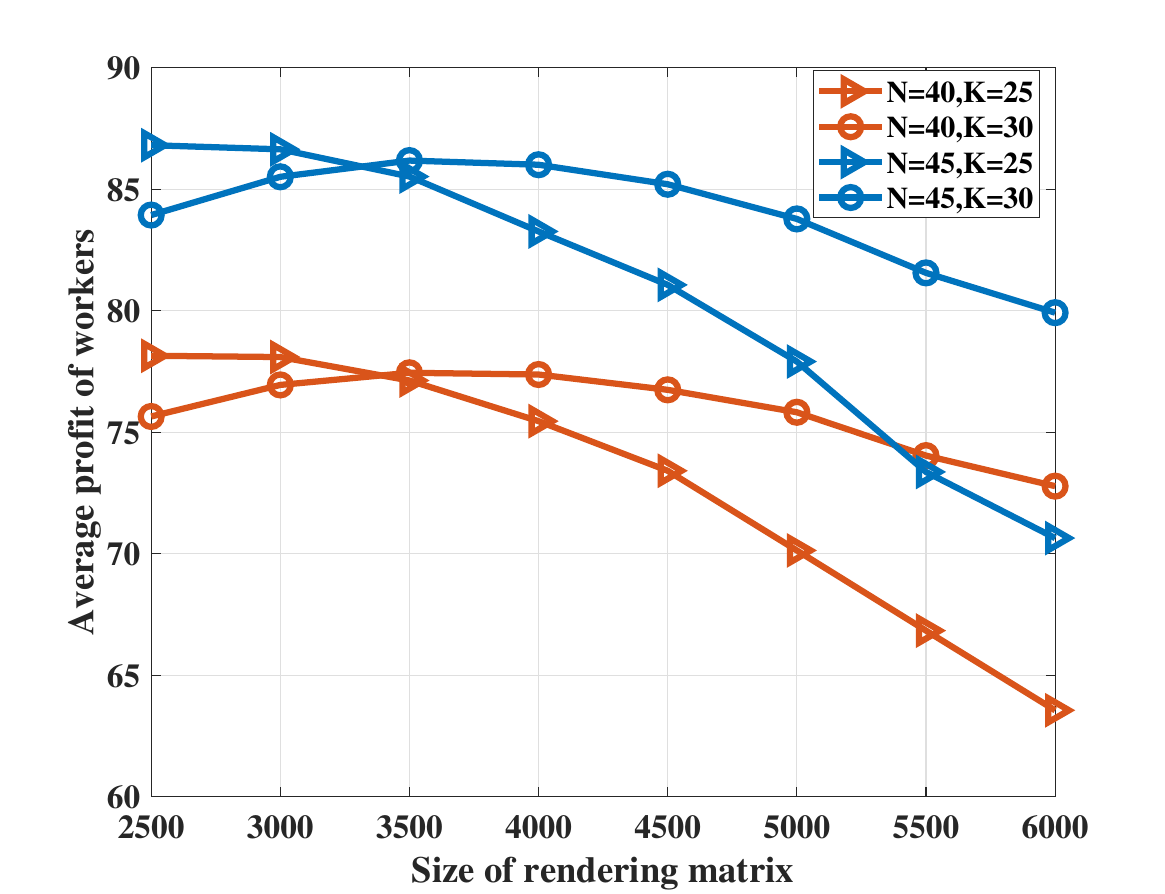}
\end{center}
\caption{Average profit of workers as function of size of rendering matrix.}
\label{Fig10}
\end{figure}
\begin{figure}[t]
	\begin{center}
		\includegraphics[width=0.4\textwidth]{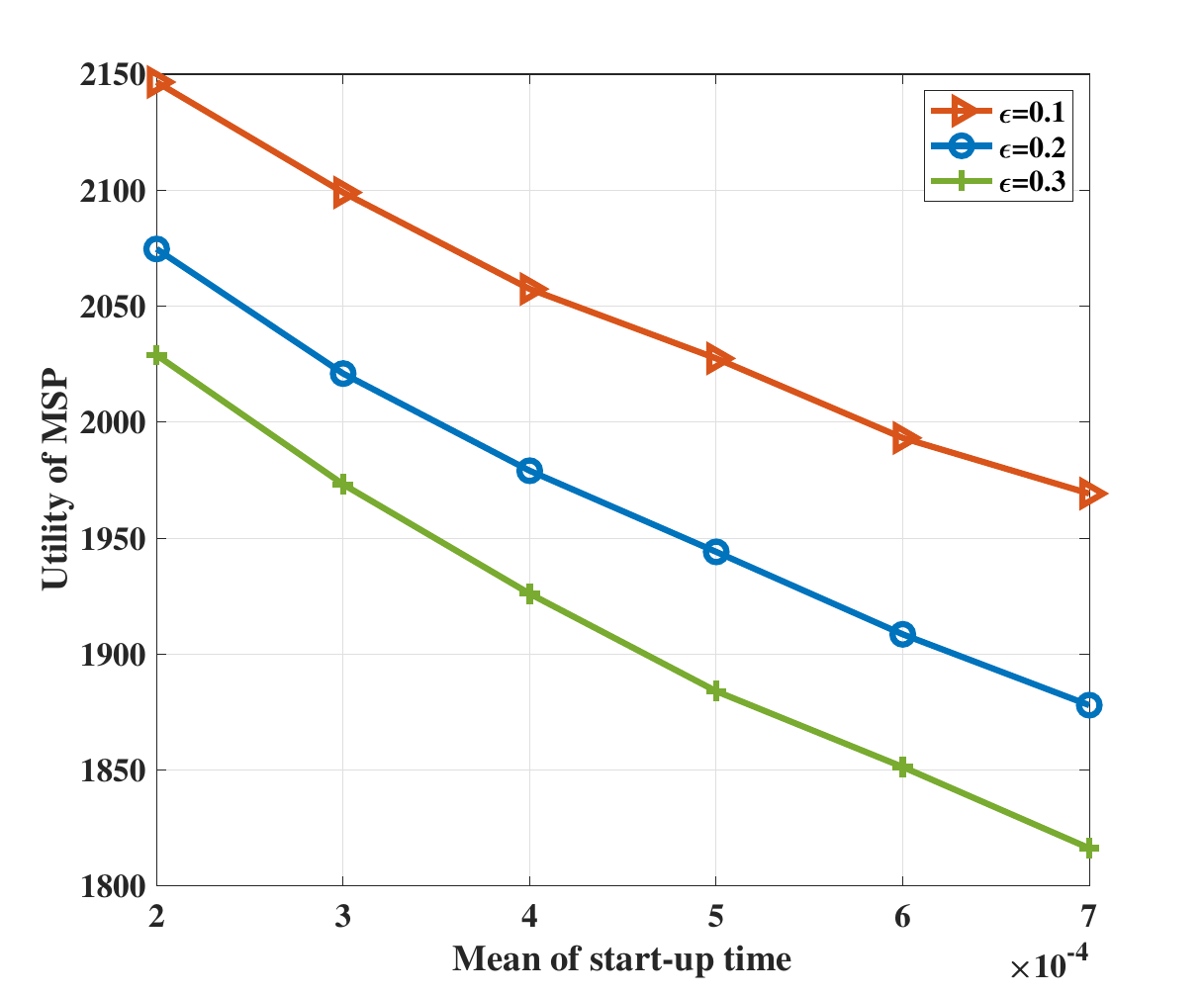}
	\end{center}
	\caption{Utility of MSP as function of the mean of start-up time.}
	\label{Fig11}
\end{figure}
\begin{figure}[t]
	\begin{center}
		\includegraphics[width=0.4\textwidth]{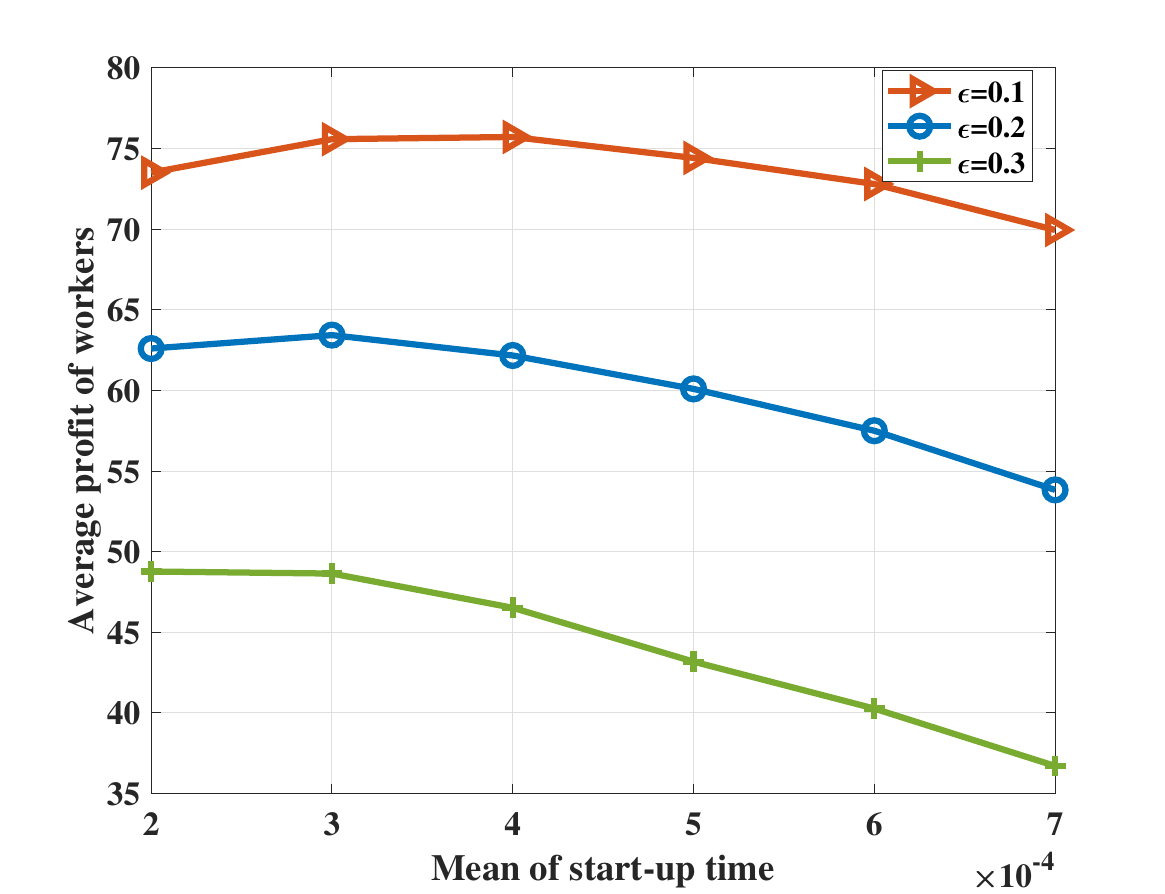}
	\end{center}
	\caption{Average profit of workers as function of the mean of start-up time.}
	\label{Fig12}
\end{figure}

In this section, according to the reliable worker selection results obtained from coalition formation game, we analyze the numerical results about Stackelberg game-based CDC incentive mechanism in the vehicular metaverse. We assume that the strat-up time of workers follows the normal distribution.
Figure 7 and Figure 8 show the effects of the size of rendering matrix on the reward given by the MSP and utility of the MSP, respectively. From Fig. 7, the reward given by the MSP increases with the size of rendering matrix. When the number of returned results $K$ is fixed, a higher value of $N$ results in higher reward given by the MSP. When $N$ is fixed, the reward decreases with the growth of $K$. As the increase of $K$ means that the MSP needs to reward more workers, which makes the reward reduced.
From Fig. 8, the utility of the MSP increases slightly with the size of rendering matrix. A higher value of $N$ results in higher utility for the MSP, which means that more workers' participation is beneficial to the MSP. However, a higher value of $K$ makes the utility of the MSP reduced. This indicates that the MSP can achieve higher utility with the CDC scheme, as the MSP obtains the final computing result when receiving the computing results from $K$ workers, and $K<N$.

Figure 9 and Figure 10 show the effects of the size of rendering matrix on the selected workers' average computation speed and average profit, respectively. From Fig. 9, the average computation speed of workers increases with the size of rendering matrix. When $K$ is fixed, a higher value of $N$ results in higher average computation speed, as the workers need to increase their speed to obtain the competition reward. When $N$ is fixed, a higher value of $K$ results in lower average computation speed. This is because the increase of $K$ reduces the competitiveness among workers. 
From Fig. 10, with the growth of the size of rendering matrix, the average profit of selected workers first increases to a peak point and then decreases. This indicates that when workers face many rendering tasks with different sizes, workers can choose to execute the rendering tasks that can maximize their profits. When $K$ is fixed, the increase of $N$ makes the workers' average utility increased. This is mainly because the MSP gives more reward when $N$ increases. When the size of rendering matrix is less than 3300 for $N = 45$ and 3500 for $N = 40$, a lower value of $K$ results in higher average profit of workers. Otherwise, a higher value of $K$ results in higher average profit of workers.

Figure 11 shows the mean of workers' start-up time on the utility of the MSP. As shown in Fig. 11, when the computing cost ${\varepsilon}$ is fixed, the utility of the MSP decreases linearly with the increase of the mean of workers' start-up time. Besides, the utility of the MSP decreases with the increase of ${\varepsilon}$. Figure 12 show the mean of workers' start-up time as function of average profit of workers. From Fig. 12, with the increase of the mean of workers' start-up time, the average profit of workers first increases to a peak point and then decreases. The workers can select suitable strat-up time to execute the computing tasks from the MSP. 

\begin{figure}[t]
	\begin{center}
		\includegraphics[width=0.4\textwidth]{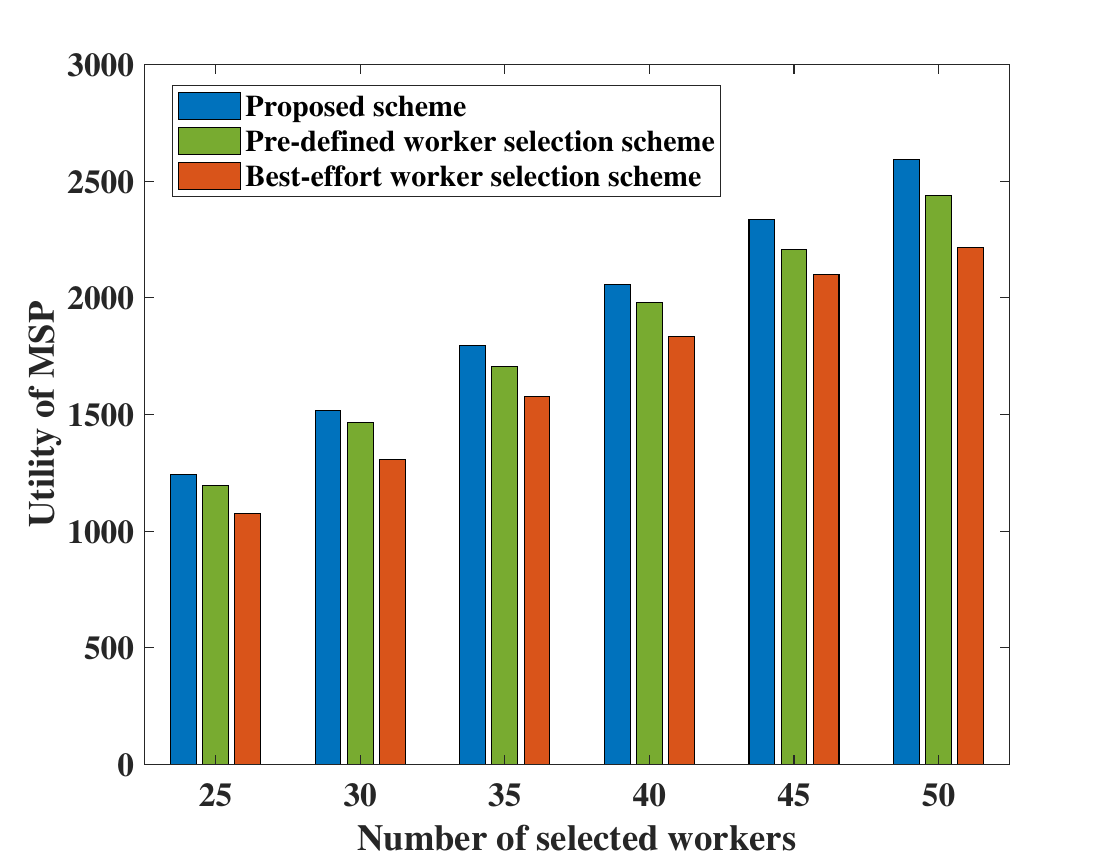}
	\end{center}
	\caption{Utility of MSP as function of number of selected workers.}
	\label{Fig13}
\end{figure}
\begin{figure}[t]
	\begin{center}
		\includegraphics[width=0.4\textwidth]{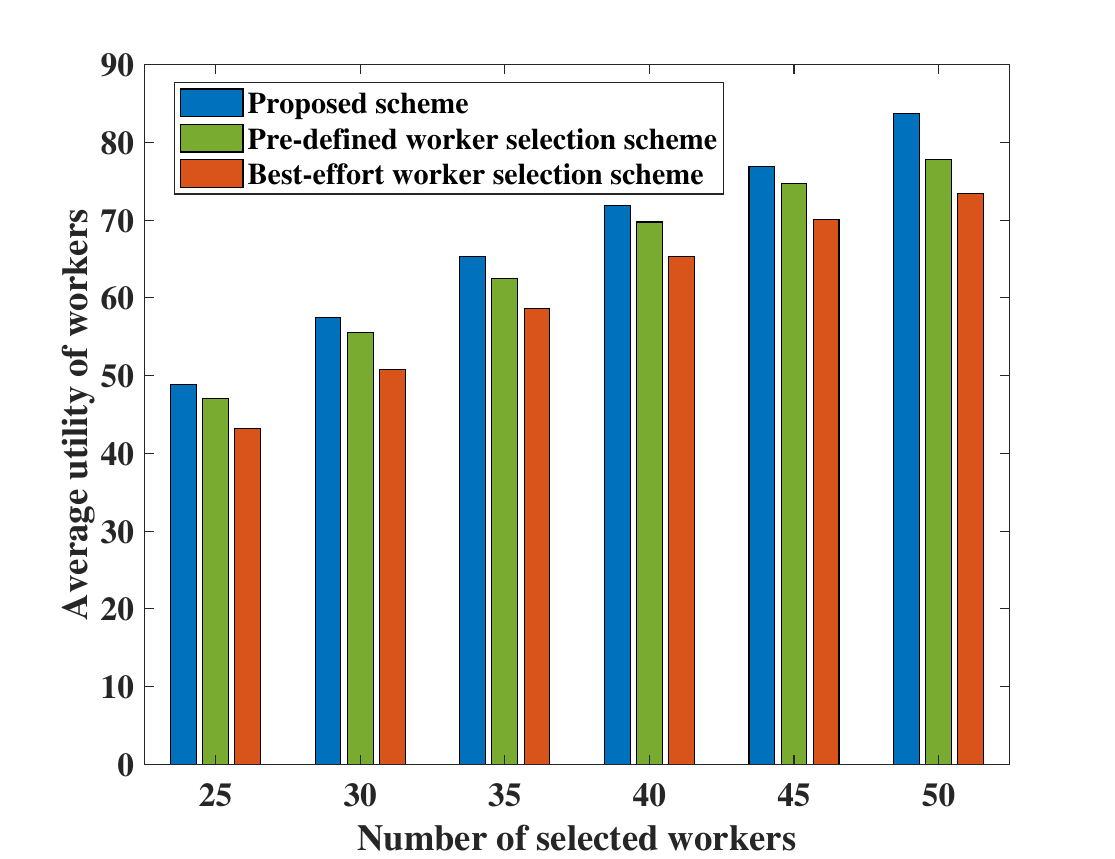}
	\end{center}
	\caption{Average profit of workers as function of number of selected workers.}
	\label{Fig14}
\end{figure}

Then, we compare the proposed scheme with the pre-defined worker selection scheme and best-effort worker selection scheme. In the pre-defined worker selection scheme, we pre-select a fixed number of workers based on certain criteria, such as their similarity \cite{comparison_scheme1}. 
In the best-effort worker selection scheme that is similar to \cite{comparison_scheme2}, any workers available will be allowed to join the CDC progress. 
When the reputation values of workers is lower than the reputation threshold $T_{th}^{com}$, the repuatation utility of the worker is expressed as $h\left( {T_{i \to w}^{com}} \right) = \alpha {e^{\left( {T_{i \to w}^{com} - T_{th}^{com}} \right)}}$ \cite{b36}. Figure 13 shows the utility of the MSP as a function of $N$ under different schemes. From Fig. 13, the utility of MSP increases with $N$ under all schemes. The utility of the MSP under the proposed reliable CDC scheme is higher than those under the pre-defined worker selection scheme and best-effort worker selection scheme. The utility of the MSP has been increased by 17\%  compared with the best-effort worker selection scheme. 
Figure 14 shows the average profit of selected workers as a function of $N$ under different schemes. From Fig. 14, the average profit of workers increases with the increase of $N$ under all schemes. The average profit of workers under the proposed reliable CDC scheme is higher than those under the pre-defined worker selection scheme and best-effort worker selection scheme. The average profit of workers has been increased by 14\% compared with the best-effort worker selection scheme.

\section{CONCLUSION}
In this paper,  a distributed computing framework is proposed for the vehicular metaverse based on CDC and blockchain. 
The subjective logical model is used to compute the reputation values of vehicles. 
A hierarchical game-theoretic CDC framework is proposed for the vehicular metaverse, the coalition formation game is combined with the reputation metric in the upper layer to select reliable workers, and the Stackelberg game is designed in the lower layer to incentivize workers to join the CDC rendering tasks. Finally, the hierarchical game-based CDC reliable worker incentive mechanism is simulated and analyzed. Simulation results indicate that the proposed reliable CDC scheme is resistant to the malicious workers and is suitable for the decentralized CDC in the vehicular metaverse. The utility of the MSP has been increased by 17\%, and the average profit of workers has been increased by 14\% compared with the best-effort worker selection scheme. In future works, specific consensus mechanisms of blockchain for the vehicular metaverse might be studied.



\end{document}